\documentclass[fleqn,usenatbib]{mnras}

\usepackage{newtxtext,newtxmath}

\usepackage[T1]{fontenc}
\usepackage{ae,aecompl}

\usepackage{graphicx}	
\usepackage{amsmath}	
\usepackage{amssymb}	
\usepackage{multicol}
\usepackage{bm}
\usepackage{booktabs}

\newcommand{\CIV}{C\textsc{iv}$\lambda \lambda 1548,1550$ }
\newcommand{\HeII}{He\textsc{ii}$\lambda1640$ }
\newcommand{\secpoint}{\mbox{$''\mskip-7.6mu.\,$}}
\newcommand{\angstrom}{\mbox{\normalfont\AA}}
\newcommand{\high}{\textit{high} }
\newcommand{\low}{\textit{low} }
\newcommand{\msun}{M_{\odot}}

\title[Massive Stars and Ionized Gas at High Redshift]{The MOSDEF-LRIS Survey: The Connection Between Massive Stars and Ionized Gas in Individual Galaxies at $z\sim2$ $^{1}$}

\author[M. W. Topping et al.]{Michael W. Topping,$^{2}$\thanks{E-mail: mtopping@astro.ucla.edu}
Alice E. Shapley,$^{2}$
Naveen A. Reddy,$^{3}$
Ryan L. Sanders,$^{4}$\newauthor
Alison L. Coil,$^{5}$
Mariska Kriek,$^{6}$
Bahram Mobasher,$^{3}$	
Brian Siana$^{3}$
\\
$^{1}$Based on data obtained at the W.M. Keck Observatory, which is operated as a scientific partnership among the California Institute of Technology, \\ the University of California,  and the National Aeronautics and Space Administration, and was made possible by the generous financial support \\ of the W.M. Keck Foundation.\\
$^{2}$Department of Physics and Astronomy, University of California, Los Angeles, 430 Portola Plaza, Los Angeles, CA 90095, USA\\
$^{3}$Department of Physics and Astronomy, University of California, Riverside, 900 University Avenue, Riverside, CA 92521, USA\\
$^{4}$Department of Physics, University of California, Davis, 1 Shields Avenue, Davis, CA 95616, USA\\
$^{5}$Center for Astrophysics and Space Sciences, Department of Physics, University of California, San Diego, 9500 Gilman Drive., La Jolla, CA 92093, USA\\
$^{6}$Astronomy Department, University of California at Berkeley, Berkeley, CA 94720, USA
}

\begin{document}
\label{firstpage}
\pagerange{\pageref{firstpage}--\pageref{lastpage}}
\maketitle

\begin{abstract}

We present constraints on the massive star and ionized gas properties for a
sample of 62 star-forming galaxies at $z\sim2.3$. Using BPASS stellar population models,
we fit the rest-UV spectra of galaxies in our sample to estimate
age and stellar metallicity which, in turn, determine the ionizing spectrum. In addition to the median properties of well-defined subsets of our sample,  we derive
the ages and stellar metallicities for 30 high-SNR individual galaxies  --
the largest sample of individual galaxies at high redshift with such
measurements.  Most galaxies in this high-SNR subsample have stellar metallicities of
$0.001<Z_*<0.004$. We then use Cloudy+BPASS  photoionization models
to match observed rest-optical line ratios and infer nebular properties.  Our high-SNR
subsample is characterized by a median ionization parameter and oxygen
abundance, respectively, of $\log(U)_{\textrm{med}}=-2.98\pm0.25$ and
$12+\log(\textrm{O/H})_{\textrm{med}}=8.48\pm0.11$. Accordingly,
we find that all galaxies in our sample show evidence for $\alpha$-enhancement.
In addition,  based on inferred $\log(U)$ and $12+\log(\textrm{O/H})$ values,
we find that the local relationship between ionization parameter
and metallicity applies at $z\sim2$.
Finally, we find that the high-redshift galaxies most offset from the local
excitation sequence in the BPT diagram are
the most $\alpha$-enhanced. This trend suggests that $\alpha$-enhancement
resulting in a harder ionizing spectrum at fixed oxygen abundance is a
significant driver of the high-redshift galaxy offset on the BPT diagram
relative to local systems.  The ubiquity of $\alpha$-enhancement among
$z\sim2.3$ star-forming galaxies indicates important differences
between high-redshift and local galaxies that must be accounted for in order to
derive physical properties at high redshift.

\end{abstract}

\begin{keywords}
galaxies: evolution -- galaxies: ISM -- galaxies: high-redshift
\end{keywords}

\section{Introduction} 
\label{sec:intro}




\begin{figure*}
    \centering
    \includegraphics[width=1.0\linewidth]{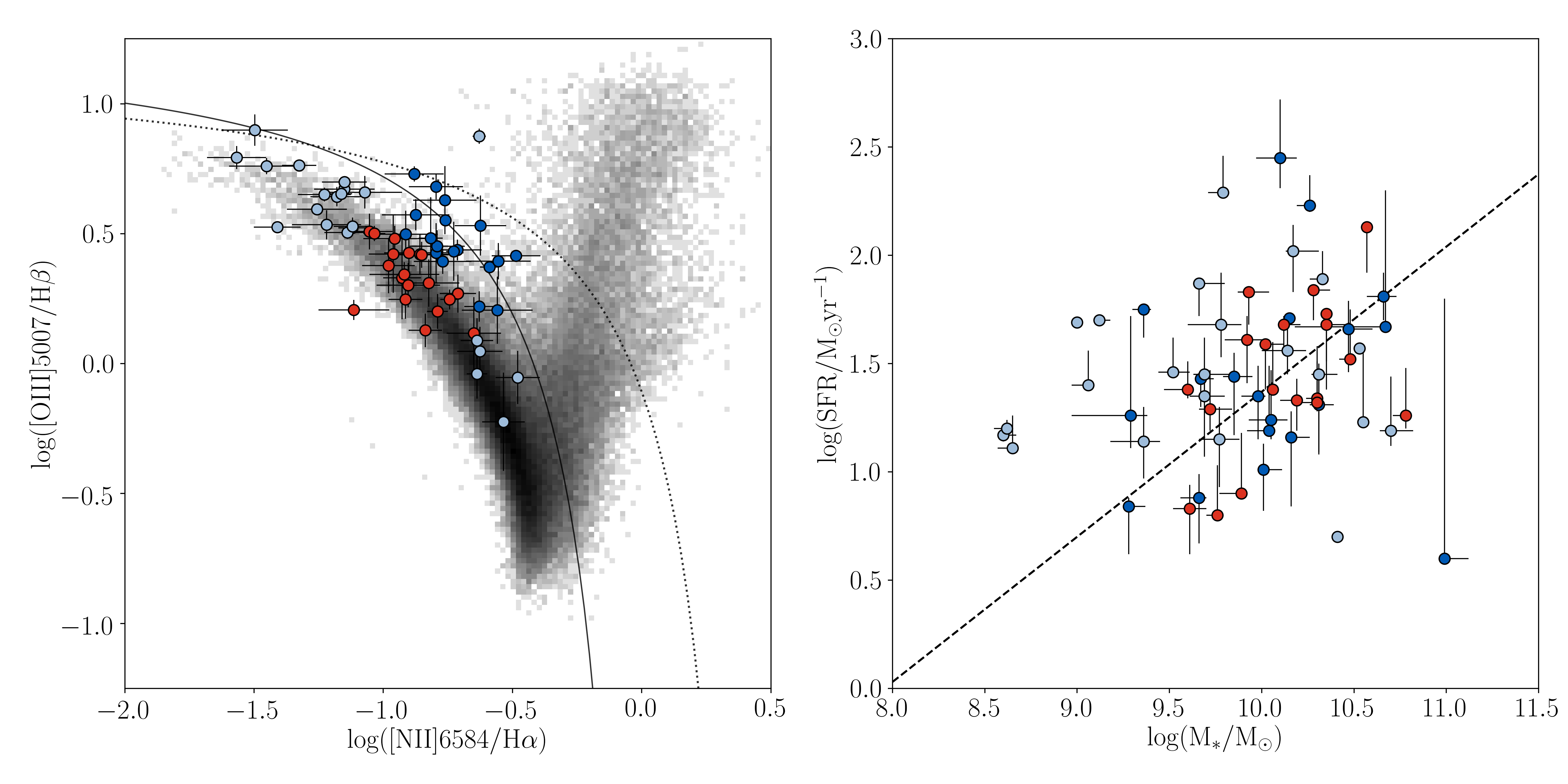}
    \caption{Properties of the LRIS-BPT sample.  Left: Location of the LRIS-BPT sample on the [OIII]$\lambda 5007$/H$\beta$ vs, [NII]$\lambda 6584$/H$\alpha$ BPT diagram.  The grey histogram shows the location of SDSS galaxies \citep[grey;][]{Abazajian2009}. Blue and red points indicate galaxies included, respectively, in the \textit{high} and \textit{low} composite spectra described in \citet{Topping2020}.   For reference, the `maximum starburst' model of \citet{Kewley2001} (dotted curve) and star-formation/AGN boundary from \citet{Kauffmann2003} (solid curve) are plotted.  Right: SFR calculated from the dust-corrected Balmer lines vs. $\textrm{M}_*$ for all objects with LRIS redshifts at $2.0\le z \le 2.7$. As in the left panel, galaxies comprising either the \high or \low stack are colored blue and red respectively.  The dashed line shows the SFR-M$_*$ relation of $z\sim2.3$ MOSDEF galaxies \citep{Sanders2018}.}
    \label{fig:lrisbptsample}
\end{figure*}

Studies of large numbers of high-redshift galaxies in the rest-optical have provided a wealth of information about the physical conditions of their interstellar medium (ISM). In the local universe, measurements of optical emission lines reveal that star-forming galaxies follow a tight sequence of simultaneously increasing [NII]$\lambda 6584$/H$\alpha$ and decreasing [OIII]$\lambda 5007$/H$\beta$ \citep[e.g.,][]{Veilleux1987,Kauffmann2003}. The shape and location of the sequence in this ``BPT" diagram \citep{Baldwin1981} reflects the physical conditions of ionized gas in the ISM of galaxies (e.g., chemical abundances, density, ionization mechanism, etc.).  Analogous observations of galaxies at high redshift expose a similar sequence, but offset toward higher [OIII]$\lambda 5007$/H$\beta$ and [NII]$\lambda 6584$/H$\alpha$ on the BPT diagram relative to local galaxies \citep{Shapley2005, Erb2006, Liu2008}.

Many properties of galaxies at high redshift may be responsible for this observed difference on the BPT diagram, including higher ionization parameters at fixed nebular metallicity, harder ionizing spectra at fixed nebular metallicity, higher densities, variations in gas-phase abundance patterns, and enhanced contributions from AGNs and shocks at high redshift \citep[see e.g.,][ for a review]{Kewley2013}.  Initial results from the MOSFIRE Deep Evolution Field \citep[MOSDEF;][]{Kriek2015} survey, suggested that the observed offset is primarily caused by an enhanced N/O ratio at fixed oxygen abundance, in addition to higher physical densities in high-redshift galaxies compared to local systems \citep{Masters2014,Shapley2015,Sanders2016a}.  Results from the Keck Baryonic Structure Survey \citep[KBSS;][]{Steidel2014} instead suggested that the offset is driven primarily by a harder intrinsic ionizing spectrum at fixed oxygen abundance \citep{Steidel2016, Strom2017}.  Updated results from the MOSDEF survey support the explanation of a harder ionizing spectrum at fixed oxygen abundance (\citealt{Sanders2020}; \citealt{Shapley2019}; \citealt{Sanders2019}; \citealt{Runco2020}). 

Furthermore, a harder stellar ionizing spectrum at fixed oxygen abundance (i.e., nebular metallicity, which traces $\rm \alpha/H$) arises naturally in the case of $\alpha$ enhancement of the most massive stars, which produce the bulk of the ionizing radiation in star-forming galaxies. For a given oxygen abundance, such $\alpha$-enhanced stars will have lower Fe/H than their local counterparts with a solar abundance pattern. A lower Fe/H corresponds to a harder ionizing spectrum. The $\alpha$-enhancement described above is expected in high-redshift galaxies due to their young ages and rising star-formation histories \citep[e.g., based on SED-fit ages in previous papers, as well as preference for rising SFHs in high-redshift galaxies][]{Papovich2011, Reddy2012b} resulting in enrichment primarily from Type II supernovae \citep{Steidel2016, Matthee2018}.

The rest-optical emission lines observed in high-redshift star-forming galaxies are strongly affected by the intrinsic ionizing spectrum primarily produced by the most massive stars. Several properties of the massive stars modulate the production of ionizing photons, including stellar metallicity, IMF, and stellar binarity \citep{Topping2015, Steidel2016, Stanway2016, Stanway2018}. In addition to controlling the ionizing spectrum, the formation of massive stars is regulated by gas accretion onto galaxies, and in turn regulates the resulting chemical enrichment of the ISM through the deposition of metals by supernova explosions and stellar winds, and the ejection of metals through star-formation-driven galaxy-scale outflows. \citet{Strom2018} investigated the relationship between properties of the ionized ISM and factors contributing to the excitation state (including stellar metallicity) for a sample of $z\sim2$ star-forming galaxies, finding that typically, $z\sim2$ galaxies have lower stellar metallicity compared to their nebular metallicity. This investigation relied purely on strong rest-optical emission lines.  However, the emission-line spectrum depends on a host of physical properties in addition to the ionizing spectral shape and is thus a highly indirect tracer of the intrinsic ionizing spectrum.  In particular, breaking the degeneracy between the ionization parameter and the ionizing spectral shape is challenging when only the strongest rest-optical emission-line ratios are available.  \citet{Sanders2020} used a small sample of star-forming galaxies at $z\sim 1.5-3.5$ with direct-method nebular metallicities to connect factors affecting the ionizing spectrum with properties of the ISM. The tight constraints on stellar metallicity facilitate breaking the degeneracy between ionization parameter and hardness of the ionizing spectrum. Another avenue for breaking this degeneracy is directly constraining the ionizing spectrum from massive stars using rest-UV spectra of these high-redshift galaxies.

While directly observing the ionizing spectrum within high-redshift galaxies is extremely challenging, information about the massive star population can be determined based on the non-ionizing rest-UV stellar spectrum that is far easier to observe. Specifically, features including the \CIV and \HeII stellar wind lines, and a multitude of stellar photospheric features are sensitive to the properties of massive stars \citep{Leitherer2001, Crowther2006, Rix2004}. For example, \citet{Halliday2008} utilized Fe\textsc{iii} absorption lines to measure a sub-solar stellar metallicity for a composite spectrum of $z\sim2$ galaxies.  \citet{Sommariva2012} tested the ability of additional photospheric absorption line indicators to accurately determine the stellar metallicity of the massive star population of high-redshift galaxies. Recently, instead of using integrated regions of the rest-UV spectrum, \citet{Cullen2019} instead fit stellar population models to the full rest-UV spectrum of multiple composite spectra to investigate galaxy properties across $2.5< z < 5.0$ and a stellar mass range of $8.5 < \log(M_*/\rm M_{\odot}) < 10.2$.

Crucially, recent studies have concurrently studied the production of the ionizing spectrum with the rest-optical properties of high-redshift galaxies by utilizing combined rest-UV and rest-optical spectroscopy \citep{Steidel2016, Chisholm2019, Topping2020}.  \citet{Steidel2016} constructed a composite spectrum of 30  $z\sim2.4$ star-forming galaxies from KBSS, and found that their rest-UV properties could only be reproduced by stellar population models that include binary stars, have a low stellar metallicity ($Z_*/Z_{\odot}\sim0.1$), and a different, higher, nebular metallicity ($Z_{\textrm{neb}}/Z_{\odot}\sim0.5$). By analyzing a single composite rest-UV spectrum, \citet{Steidel2016} only probed average properties of their high-redshift galaxy sample. With single rest-UV and rest-optical composite spectra it is not possible to probe the rest-UV spectral properties as a function of the location in the BPT diagram. In contrast, \citet{Chisholm2019} fit linear combinations of stellar population models to 19 individual galaxy rest-UV spectra at $z\sim2$, and determined light-weighted properties. Using these individual spectra, \citet{Chisholm2019} found using mixed age populations that galaxies in their sample had comparable stellar and nebular metallicities.  In \citet{Topping2020} we compared the properties of two composite spectra one of which included galaxies lying along the local sequence and the other including galaxies offset towards high  [NII]$\lambda 6584$/H$\alpha$ and [OIII]$\lambda 5007$/H$\beta$.  This analysis indicated that galaxies offset from the local sequence had younger ages and lower stellar metallicities on average, resulting in a harder ionizing spectrum in addition to a higher ionization parameter and a higher $\alpha$-enhancement, all of which contributed to the difference in BPT diagram location.  Intriguingly, we found that even high-redshift galaxies coincident with local star-forming sequence on the BPT diagram were $\alpha$-enhanced ($\rm O/Fe \sim 3 \textrm{ O/Fe}_{\odot}$) relative to their local counterparts , suggesting that coincidence in the BPT diagram does not necessarily imply similar ISM conditions.

In this paper, we improve the methods presented in \citet{Topping2020} by expanding the stellar population synthesis models to consider more complex star formation histories (SFHs), and including additional rest-optical emission lines to the photoionization modelling. Furthermore, we test the capability of the models to be fit to individual galaxy spectra that have lower SNR than composite spectra, and analyze a sample of $\sim30$ individual galaxies with combined rest-UV and rest-optical spectra.

The organization of this paper is as follows: Section 2 describes our observations, data reduction, and methods.  Section 3 presents the results of our analysis, and Section 4 provides a summary and discussion of our key results. Throughout this paper we assume a cosmology with $\Omega_m = 0.3$, $\Omega_{\Lambda}=0.7$, $H_0=70 \textrm{km s}^{-1}\ \textrm{Mpc}^{-1}$, and adopt solar abundances from \citet[][i.e., $Z_{\odot}=0.014$, $12+\log(\rm O/H)_{\odot}=8.69$]{Asplund2009}.

\section{Methods and Observations}
\label{sec:methods}

\subsection{Rest-Optical Spectra and the MOSDEF survey}

Our analysis utilizes rest-optical spectroscopy of galaxies from the MOSDEF survey \citep{Kriek2015} at $z\sim2.3$, observed using the Multi-Object Spectrometer for Infra-Red Exploration \citep[MOSFIRE;][]{McLean2012} over 48.5 nights during 2012--2016. This rest-optical spectroscopic sample is composed of $\sim1500$ near-infrared spectra at moderate resolution ($\textrm{R}\sim3500$) of $H$-band selected galaxies targeted to lie within three distinct redshift intervals ($1.37 \le z \le 1.70$, $2.09 \le z \le 2.61$, and $2.95 \le z \le 3.80 $). Based on the scatter between photometric and spectroscopic redshifts of the MOSDEF targets, the actual redshift ranges slightly differ from the initial target ranges. Therefore, we define the true redshift ranges as $1.40 \le z \le 1.90$, $1.90 \le z \le 2.65$, and $2.95 \le z \le 3.80 $. In addition to rest-optical spectra from the MOSDEF survey, MOSDEF targets have extensive ancillary datasets from the CANDELS \citep{Grogin2011} and 3D-HST \citep{Momcheva2016} surveys. MOSDEF spectra were used to measure fluxes and redshifts of all rest-optical emission lines detected within the Y, J, H, and K bands, the strongest of which are: [OII]$\lambda3727$, H$\beta$, [OIII]$\lambda \lambda 4959,5007$, H$\alpha$, [NII]$\lambda6584$, and [SII]$\lambda \lambda 6717,6731$.

\subsection{Rest-UV Spectra and the MOSDEF-LRIS sample}

A full description of the rest-UV data collection and reduction procedures will be described in a future work, but we provide a brief description here.  We selected a subset of MOSDEF galaxies for rest-UV spectroscopic followup using the Low-Resolution Imaging Spectrograph \citep[LRIS;][]{Oke1995}.  Target priorities were determined using the following prescription.  Highest priority was given to objects from the MOSDEF survey that had detections in all four BPT emission lines (H$\beta$, [OIII], H$\alpha$, [NII]$\lambda 6584$) with $\ge 3 \sigma$.  Then, objects were added to the sample with detections in H$\beta$, [OIII]$\lambda 5007$, and H$\alpha$ with $\ge 3 \sigma$, and an upper limit in [NII]$\lambda 6584$.  With decreasing priority, the remaining targets were selected by having a spectroscopic redshift measurement from the MOSDEF survey, objects from the MOSDEF survey without a successfully measured redshift, and objects not targeted as part of the MOSDEF survey, but that were part of the 3D-HST survey catalog \citep{Momcheva2016} with properties within the MOSDEF survey photometric redshift and apparent magnitude ranges.  In total, these targets comprise a sample of 260 galaxies.  Of those targets with spectroscopic redshifts from the MOSDEF survey, 32, 162, and 20 were in the redshift ranges $1.40 \le z \le 1.90$, $1.90 \le z \le 2.65$, and $2.95 \le z \le 3.80 $ respectively.  The remaining galaxies, with either a spectroscopic redshift not from MOSDEF, or a photometric redshift, made up 9, 31, and 6 galaxies in the redshift intervals $1.40 \le z \le 1.90$, $1.90 \le z \le 2.65$, and $2.95 \le z \le 3.80 $ respectively. For this analysis, we excluded the few objects identified as AGN based on their X-ray and rest-frame near-IR properties.

 \begin{figure}
    \centering
    \includegraphics[width=1.0\linewidth]{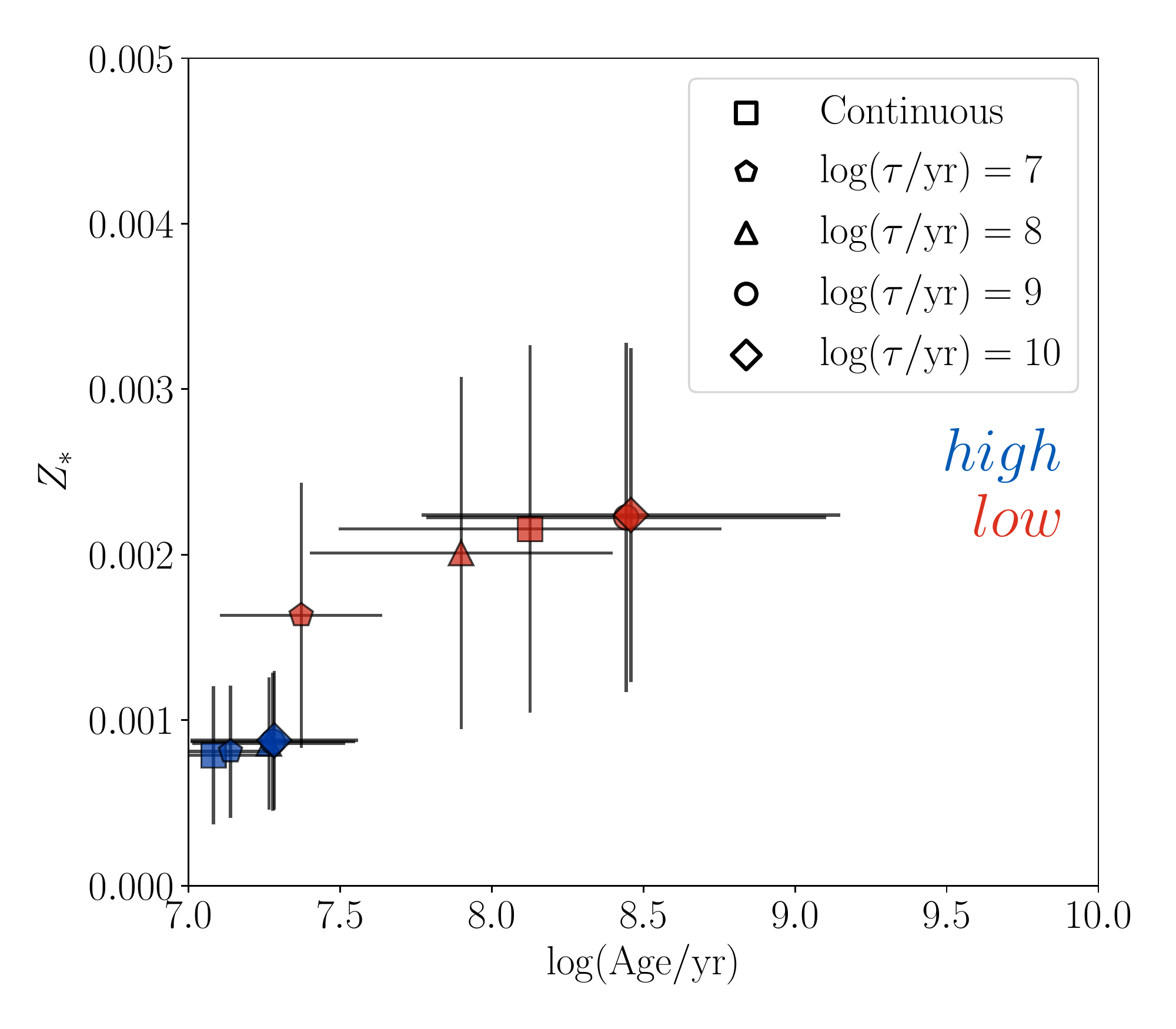}
    \caption{Best-fit stellar metallicity and age for the \textit{high} (blue) and \textit{low} (red) for five different star formation histories including a continuous SFH, and four realizations of the delayed-$\tau$ model, each depicted by a different shape. In all cases, galaxies in the \high stack have younger ages and lower stellar metallicities compared to the \low stack.  The best-fit age and stellar metallicity increases with increasing $\tau$ when considering models with a `delayed-$\tau$' SFH.  }
    \label{fig:bptallstellarfits}
\end{figure}

Rest-UV spectra were obtained using Keck/LRIS during ten nights across five observing runs between January 2017 and June 2018. Our target sample totals 260 distinct galaxies  on 9 multi-object slit masks with $1\secpoint2$ slits in the COSMOS, AEGIS, GOODS-S, and GOODS-N fields. In order to obtain continuous wavelength coverage from the atmospheric cut-off at $3100\angstrom$ up to a median red wavelength limit of $\sim$7650$\textrm{\angstrom}$, we observed all slit masks using the d500 dichroic, the 400 lines mm$^{-1}$ grism blazed at $3400\textrm{\angstrom}$ on the blue side, and the 600 lines mm$^{-1}$ grating blazed at $5000\textrm{\angstrom}$ on the red side. This setup yielded a resolution of $R\sim800$ on the blue side, and a resolution of $R\sim1300$ on the red side. The exposure times ranged from 6--11 hours on different masks, with a median exposure time of $\sim 7.5$ hours for the full sample. The data were collected with seeing ranging from $0\secpoint6$ to $1\secpoint2$ with typical values of $0\secpoint8$.

\begin{figure*}
    \centering
    \includegraphics[width=1.0\linewidth]{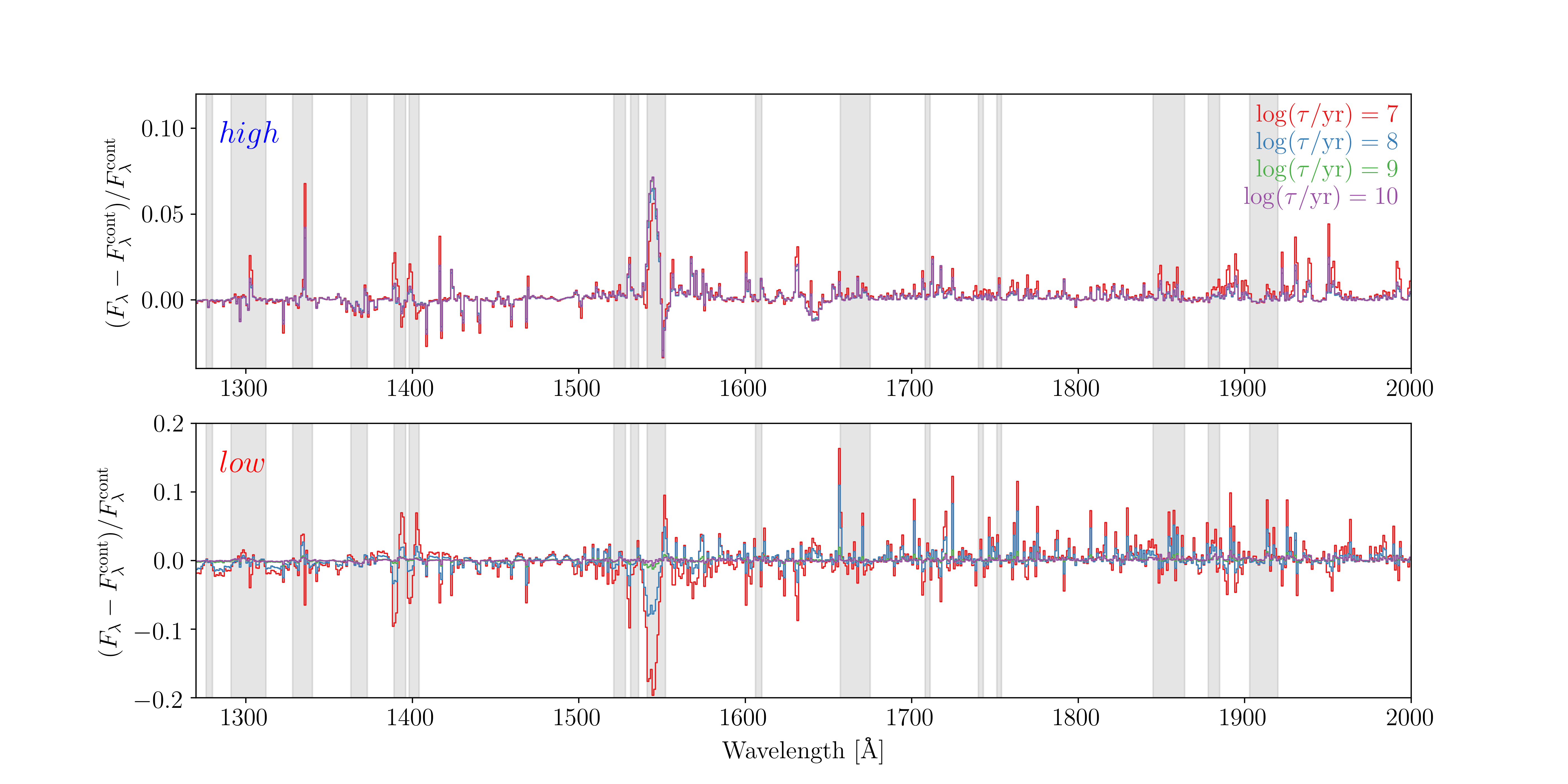}
    \caption{Fractional difference between the best-fit model spectra from models using a delayed-$\tau$ SFH compared to the model spectrum using a constant SFH.  Best-fit models fit to the \high and \low stacks are displayed in the top and bottom panels respectively.  The regions masked out using `mask1' from \citet{Steidel2016} defined to include contamination from non-stellar sources is shown in grey.  On average, the models assuming a delayed-$\tau$ SFH differ from those with a constant SFH at the few percent level.}
    \label{fig:bestfitspectra}
\end{figure*}

We reduced the red- and blue-side data from LRIS using custom \texttt{iraf}, \texttt{idl}, and \texttt{python} scripts. First, we fit polynomials to the edges of each 2D slitlet, and transformed it to be rectilinear. The subsequent steps required slightly different treatment for the red and blue spectral images.  We flat fielded each image using twilight sky flats for the blue side, and dome flats for the red side images.  Then we cut out each slitlet for each object in every flat-fielded exposure.  Following this step, the blue-side slitlets were cleaned of cosmic rays, and background subtracted. These images were registered and median combined to create a stacked two-dimensional spectrum.  In order to prevent over-estimation of the background due to the target, we measured the trace of each object in the stacked two-dimensional spectrum, and masked it out for a second-pass background subtraction \citep{Shapley2006}. As the red-side images were more significantly affected by cosmic rays, we reduced the red side using slightly different methods.   For the red-side slitlets, we constructed a stacked two-dimensional spectrum by first registering and median combining the images using minmax rejection to remove cosmic rays. We used this stacked image to measure the object traces in each slitlet. We then recomputed the background subtraction in the individual images with the object traces masked out, as the stacked image is too contaminated by sky lines to achieve a good background subtraction.  After the second pass background subtraction, the individual red-side slitlets were combined to create the final stacked image.

We extracted the 1D spectrum of each object from the red and blue side stacked slitlets.  We calculated the wavelength solution by fitting a 4th-order polynomial to the red and blue arc lamp spectra, which resulted in residuals of $\sim0.035\textrm{\angstrom}$ and $\sim0.3\textrm{\angstrom}$ respectively. We repeated this step on a set of frames that had not had sky lines removed.  Using the resulting sky spectra, we measured the centroid of several sky lines and shifted the wavelength solution zeropoint until the sky lines aligned with their known wavelengths. This shift typically had a magnitude of $\sim 4 \angstrom$ throughout the sample. We applied an initial flux calibration based on spectrophotometric standard star observations obtained during each observing run. We checked the flux calibration by comparing spectrophotometric measurements calculated from our objects to measurements in the 3D-HST photometric catalog, and applied a multiplicative factor to correct our calibration. Throughout our sample, this factor had median magnitude of $\sim30\%$. Following the final flux calibration, we ensured that continuum levels on either side of the dichroic at $5000\angstrom$ were consistent.A small number of galaxies in our sample required additional wavelength-dependent continuum correction near ($\lesssim50 \rm \angstrom$) the dichroic cutoff.  To apply this correction we defined a target continuum, $F_{\rm cont}^{\rm targ}$ using regions of the spectrum just outside of the area that required correction.  We then fit a linear function to the part of the observed spectrum to be corrected, $F_{\rm cont}^{\rm obs}$.  Finally, we multiplied this piece of the spectrum by $F_{\rm cont}^{\rm targ}/F_{\rm cont}^{\rm obs}$ to recover the correct shape of the spectrum near the dichroic cutoff.

While the full sample described above consists of $260$ galaxies across three distinct redshift intervals ($1.40 \le z \le 1.90$, $1.90 \le z \le 2.65$, and $2.95 \le z \le 3.80 $), we focus on a subset of this sample composed of galaxies in the central redshift window that have detections in four primary BPT lines (H$\beta$, [OIII]$\lambda 5007$, H$\alpha$, [NII]$\lambda6584$) at $\ge3\sigma$ from the MOSDEF survey. This `LRIS-BPT' sample comprises 62 galaxies, each of which has a systemic redshift measured from rest-optical emission lines.

\begin{table*}
\begin{center}
\begin{tabular}{ll}
\toprule

$\log(\textrm{Age/yr})$ &$7.0, 7.3, 7.4, 7.5, 7.7, 8.0, 8.5, 9.0, 9.6$\\
 $Z_*$ &$10^{-5}, 10^{-4}, 0.0001,0.001, 0.002, 0.003, 0.004, 0.006, 0.008, 0.01, 0.014, 0.02, 0.03, 0.04$\\
  $\log(Z_{\textrm{neb}}/Z_{\odot})$ &-1.3, -1.0, -0.8, -0.6, -0.5, -0.4, -0.3, -0.2, -0.1, 0.0, 0.1, 0.2\\
   $\log(U)$ &-3.6, -3.4, -3.2, -3.0, -2.8, -2.6, -2.4, -2.2, -2.0, -1.8, -1.6, -1.4\\
 \bottomrule
 \end{tabular}
 \end{center}
 \caption{Summary of model grid parameters.  The age and stellar metallicity values correspond to BPASS models we fit to our observed spectra. For each combination of age and stellar metallicity, we computed a set of photoionzation models with the listed nebular metallicity and ionization parameter values.  On this scale $\log(Z_{\textrm{neb}}/Z_{\odot})=0$ corresponds to $\rm 12+\log(\rm O/H)=8.69$} 
\label{table:cloudygrid}
\end{table*}

Figure~\ref{fig:lrisbptsample} shows the galaxy properties of objects in the LRIS-BPT sample. \citet{Topping2020} constructed composite spectra from two subsets of the LRIS-BPT sample, each subset being defined based on their location on the BPT diagram (Figure~\ref{fig:lrisbptsample}, left).  These two subsets, referred to as the \textit{high} and \textit{low} samples, comprise galaxies that are offset from the local BPT sequence, and those that lie along the sequence respectively.  Figure~\ref{fig:lrisbptsample} (right) shows the SFR measured from dust-corrected Balmer lines and stellar mass for the galaxies in our sample. The SFR and stellar mass of this sample are characterized by a median and inner $68^{th}$ percentile $\log(\textrm{SFR}/(\rm M_{\odot}/\textrm{ yr}))=1.53\pm0.44$ and $\log(\textrm{M}_*/\textrm{M}_{\odot})=10.02\pm0.52$ respectively. These median values are consistent with those of the sample of galaxies comprising the central redshift range ($1.90 \le z \le 2.65$) of the full MOSDEF survey, which are characterized by median and inner $68^{th}$ percentile of $\log(\textrm{ SFR}/(\rm M_{\odot}/\textrm{yr}))=1.36\pm0.50$ and $\log(\textrm{ M}_*/\textrm{M}_{\odot})=9.93\pm0.60$. These comparisons of SFR and stellar mass suggest that our LRIS-BPT sample is an unbiased subset of the full $z\sim2$ MOSDEF sample.

\subsection{Stellar Population Synthesis and Photoionization Models}

For this analysis, we used the version 2.2.1 Binary Population and Spectral Synthesis (BPASS) stellar population models to interpret our observed rest-UV galaxy spectra \citep{Eldridge2017, Stanway2018}. Notably, these stellar population models incorporate the effects of stellar rotation, quasi-homogeneous evolution, stellar winds, and binary stars.  These effects can have a substantial effect on the spectrum of a model stellar population, and in particular, the EUV spectrum produced by massive, short-lived stars. The BPASS models are evaluated with multiple Initial Mass Functions (IMFs), including the \citet{Chabrier2003} IMF, and IMFs with high-mass ($M\ge0.5\rm M_{\odot}$) slopes of $\alpha=-2.00$, $-2.35$, and $-2.70$.  In addition, the models using each IMF were formulated with a high-mass cutoff of $100\msun$ and $300\msun$. For this analysis, we only considered models that assume a \citet{Chabrier2003} IMF, and a high-mass cutoff of $100\msun$. Finally, the BPASS models have been computed with ages between $\log(\textrm{Age/yr})=6.0-11.0$ in increments of $0.1$ dex, and stellar metallicities of $Z_*=10^{-5}-0.04$.  While we considered all available stellar metallicities in our analysis, we restricted the ages to $\log(\textrm{Age/yr})=7.0-9.6$.  At ages younger than $\log(\textrm{Age/yr})=7.0$ we would be probing timescales shorter than the dynamical timescale of the galaxies, and therefore could not accurately attribute physical properties to the entire galaxy simultaneously \citep{Reddy2012b, Price2020}. This minimum age restriction is further motivated by the H$\alpha$ EW distribution of our sample, which is characterized by a median value of $150\angstrom$, consistent with galaxies that have ages of $\gtrsim 10\rm Myr$ \citep{Erb2006}.  Even the most extreme galaxies in our sample have H$\alpha$ EWs of $\sim800\angstrom$, which is consistent with our range of stellar population model ages (see \citet{Reddy2018} for the H$\alpha$ EW distribution of the full MOSDEF sample). Additionally, the ionizing spectrum of stellar populations with constant star formation reaches an equilibrium at an age $\sim10^7\textrm{yr}$. Finally, at the lowest redshift galaxy in our sample, the age of the universe was $\sim4$Gyr, so including older stellar populations is not necessary.

\begin{figure*}
    \centering
    \includegraphics[width=1.0\linewidth]{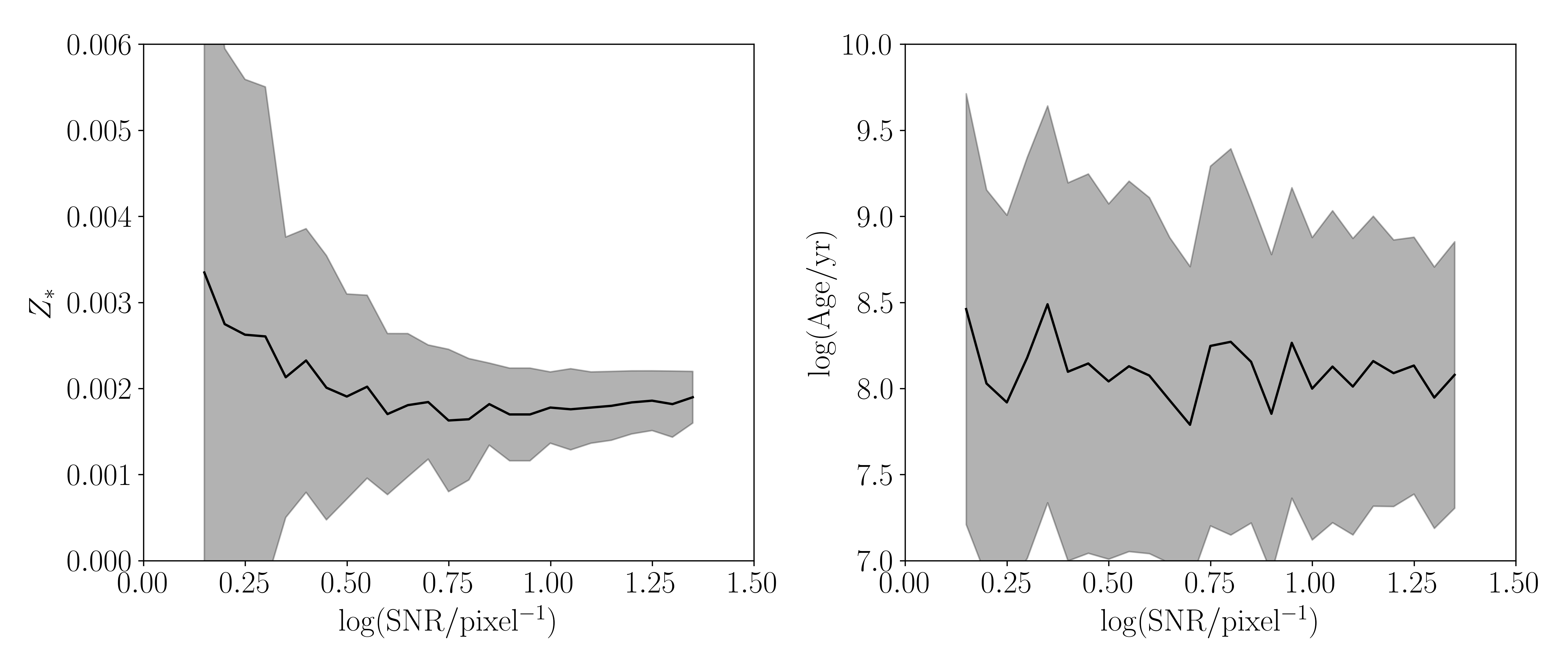}
    \caption{The best-fit stellar metallicity (left) and stellar age (right) as a function of spectra SNR created by artificially adding an increasing amount of noise to one of our composite spectra.  The best-fit stellar metallicity remains consistent to the high-SNR value in the range $\rm SNR\ per\ spectral\ resolution\ element \ge 5.6$.  The best-fit age remains consistent at all SNR values, however the $1\sigma$ uncertainties increase to the size of the parameter space ($7.0 \le \log(\textrm{Age/yr}) \le 9.6$) at low SNR. }
    \label{fig:stackscaledsnr}
\end{figure*}

We constructed stellar population models that assume different star-formation histories (SFH) by combining the BPASS models, which describe a coeval stellar population, using:
\begin{equation}
    F_{\lambda} = \Psi_{t_0} f(\lambda)_{t_0}\Delta t_0 + \sum_{i=1}^{t_{\textrm{max}}} \Psi_{t_i} f(\lambda)_{t_{\textrm{max}}-t_i}(t_i-t_{i-1}),
\end{equation}
where $t_{\textrm{max}}$ is the age of the population, $\Delta t_0$ is the time between the age of the first model (i.e., $10^6$yr) and the onset of star formation, $\Psi_{t_0}$ is the SFR at the time of the first model, $f(\lambda)_{t_0}$ is the luminosity density of the model spectrum per unit stellar mass at the time of the first model,  $\Psi_{t_i}$ is the star-formation rate of the population at time $t_i$, $f(\lambda)_{t_{\textrm{max}}-t_i}$ is the luminosity density of the model spectrum per unit stellar mass with age $t_{\textrm{max}}-t_i$ (i.e., the model that began $t_i$ years prior to the final age, $t_{\textrm{max}}$), and $(t_i-t_{i-1})$ is the time between subsequent model spectra. For the case of a constant SFH, all of the SFR weightings, $\Psi_{t_i}$, are set to unity. Each spectrum was constructed using models with the same stellar metallicity (i.e., none of our model spectra represent a combined population of multiple stellar metallicities). In addition to a constant SFH, we considered several models with a `delayed-$\tau$' SFH of the form $\textrm{SFR}\propto t\times e^{-t/\tau}$, with $\log(\tau/\textrm{yr})=7,8,9,10$.  With this set of models we covered three schematically different SFHs.  These SFHs allowed us to explore models when star formation is rising ($t<\tau$), falling($t>\tau$), and peaked($t\sim\tau$).

\begin{figure*}
    \centering
    \includegraphics[width=1.0\linewidth]{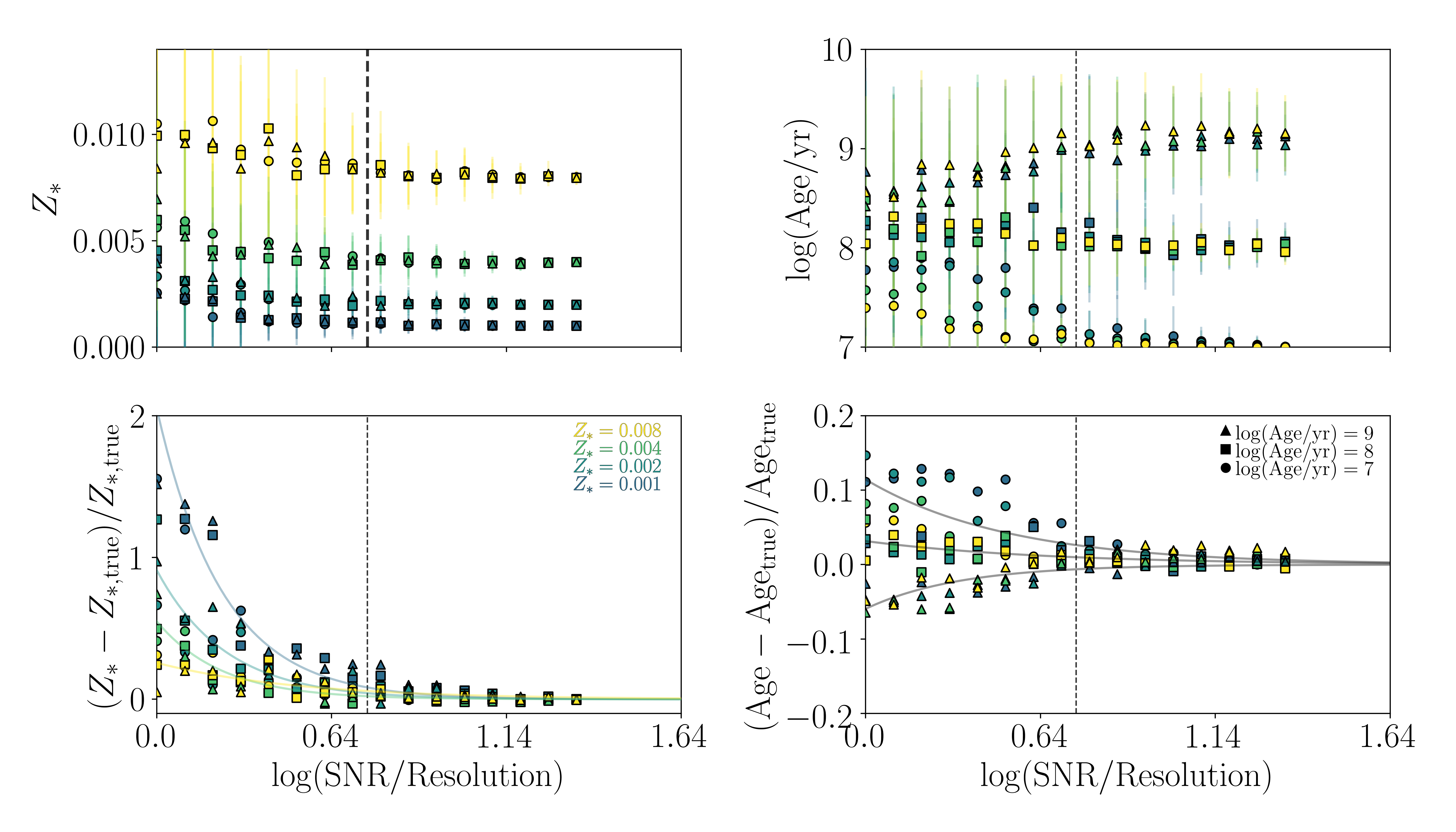}
    \caption{Best-fit stellar parameters as a function of SNR computed by adding varying amounts of noise to an array of BPASS model spectra for which the $\log(\textrm{Age/yr})$ and $Z_*$ are known. For all panels the color corresponds to the model input stellar metallicity, and the symbol depicts the input stellar age. Top Right: Best-fit stellar metallicities as a function of SNR computed for four different input values (0.001, 0.002, 0.004, 0.008) each of which has been computed at three different ages.  At high SNR/resolution element all of the models asymptote to their input values, however at SNR/resolution element $\le 5.6$ the best-fit values are biased high. Bottom Right: Same as top right but for the fractional difference between the best-fit $Z_*$ and the input $Z_*$.  At the lowest SNR, the best-fit values can be biased high by $\sim50\%-150\%$. Top Left: Best-fit stellar ages as a function of SNR for three different input ages at a range of stellar metallicities.  At the lowest SNR, the uncertainties expand to fill the parameter search range, and the best-fit values are biased toward $\log(\textrm{Age/yr})\sim8.5$. Bottom Right: Fractional difference between the best-fit age and the input age for each of our models. The best-fit results begin to significantly diverge from their input values at $<\rm SNR\ per\ spectral\ resolution\ element \sim 5.6$, indicated by the vertical dashed line. }
    \label{fig:modelscaledsnr}
\end{figure*}

We processed the stellar population model spectra using the photoionization code Cloudy \citep{Ferland2017}. Using this code, we input an ionizing spectrum from BPASS and, given a set of ISM properties, calculated expected emission line fluxes.  We compared the simulated line fluxes to the observed rest-optical emission lines of galaxies in our sample to infer properties of the ISM.  We assumed a fixed electron density of $n_e=250\textrm{  cm}^{-3}$, which is representative of the galaxies in our sample \citep{Sanders2016}. In addition, while we vary the nebular oxygen abundance, we assume solar abundance ratios for most elements. However, we adopt the $\log(\textrm{N/O})$ vs. $\log(\textrm{O/H})$ relation from \citet{Pilyugin2012}: \vspace{0.4cm}

      \ $\log(\textrm{N/O})=-1.493$ \\
     \vspace{.1cm}
      \indent \indent for $12+\log(\textrm{O/H}) < 8.14$\\
           \vspace{.1cm}
     \indent $\log(\textrm{N/O})=1.489 \times [12+\log(\textrm{O/H})] - 13.613$ \\
          \vspace{.1cm}
      \indent \indent for $12+\log(\textrm{O/H}) \ge 8.14$. 
 \vspace{0.1cm}
 
For each BPASS model, we ran a grid of Cloudy models for a range of nebular metallicity ($-1.6 \le \log(Z_{\textrm{neb}}/Z_{\odot}) \le 0.3$) and ionization parameter ($-3.6 \le \log(U) \le -1.4$).  We have made several updates to the parameters of the model grids described in \citet{Topping2020}, in order to more finely sample the parameter space in regions of interest, including amending the abundances of all elements to the values reported by \citet{Asplund2009}.  Table~\ref{table:cloudygrid} summarizes the parameters, and lists each value for which we compute a model. An additional component of the photoionization models is the nebular continuum.  The nebular continuum contributes a relatively small amount of flux to the UV spectrum, compared to the stellar component (See \citet{Steidel2016}, Figure 3). We explicitly compute the nebular continuum for BPASS models listed in Table~\ref{table:cloudygrid}, however, it changes smoothly with age, so we interpolate the nebular contribution for the remaining BPASS models.

\subsection{Composite Spectra and Fitting}
 
To compute a composite spectrum, we first interpolated each of the individual galaxy spectra onto a common wavelength grid. We chose the sampling of this common wavelength grid to be $0.8\angstrom$, which corresponds to the rest-frame sampling of our spectra at the median redshift of our sample. Then, we continuum normalized each of the individual galaxy spectra. In the process of continuum fitting, we first extracted regions of the spectra that are not contaminated by absorption lines, in the windows defined by \citet{Rix2004}. We then fit a cubic spline to the median flux values within each window to define the continuum level. Because of this approach, we did not need to consider effects that smoothly affect the continuum (e.g., reddening by dust). Following this step, at each wavelength, we median combined all spectra that had coverage at that wavelength.  Only 3/62 of the galaxies in our sample do not cover the full wavelength range that we are fitting, therefore a composite spectrum at any given wavelength is well sampled by the individual galaxy spectra it comprises.  We defined the error spectrum as the standard deviation of all contributing spectra at each wavelength.    
 
To fit the BPASS stellar population synthesis models to our individual galaxy and composite spectra, we masked out regions of the observed spectra that include components not present in the models (e.g., strong interstellar absorption lines). Then, we continuum normalized both the observed and BPASS model spectra.  We then interpolated the BPASS models onto the wavelength grid of the galaxy spectra. We did not smooth either the models or the observed spectra as their resolutions were comparable with values of $\sim1\angstrom$ in the rest-frame. Following this step, we calculated the $\chi^2$ statistic for each BPASS model in the grid, and determined which age and stellar metallicity produced the minimum $\chi^2$ value. We determined the uncertainties in these parameters by perturbing the observed spectrum and calculating which age and stellar metallicity best-fit the observed spectrum.  In the case of an individual galaxy spectrum, this perturbation simply constitutes adding in noise to each wavelength element pulled from a normal distribution with a standard deviation defined by the magnitude of the error spectrum at that wavelength element.  For a simulated composite spectrum, we selected a new sample of galaxies from the initial composite spectrum sample with replacement.  Then, each galaxy spectrum was perturbed using the method described above before being combined. After repeating this process 1000 times, we defined the best-fit value and upper and lower $1\sigma$ uncertainties as the median, $16^{\textit{th}}$ and $84^{\textit{th}}$ percentile of the distribution.

\begin{figure}
    \centering
    \includegraphics[width=1.0\linewidth]{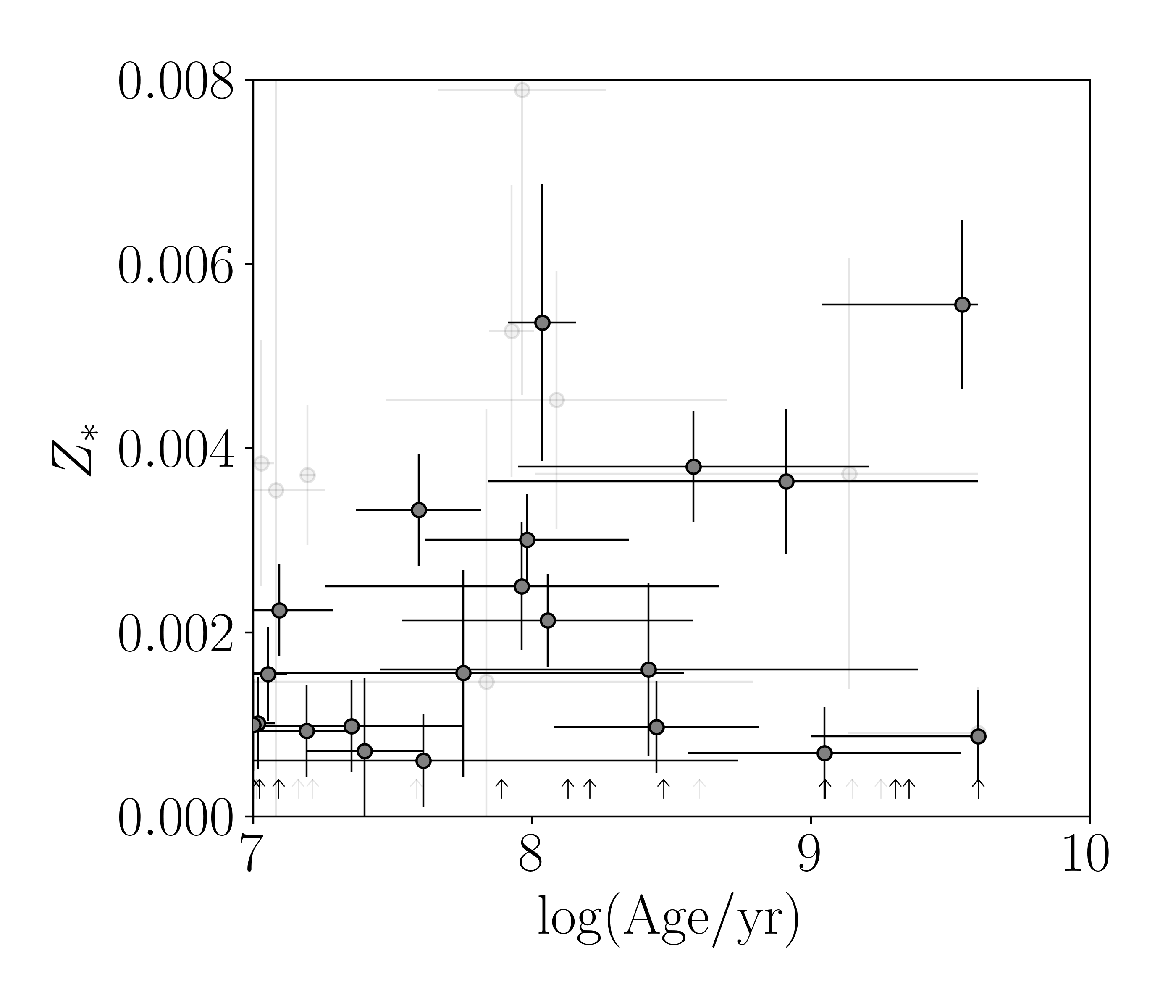}
    \caption{Best-fit age and stellar metallicity for all galaxies in the LRIS-BPT sample.  For completeness, galaxies with $\rm SNR\ per\ spectral\ resolution\ element \le 5.6$ are displayed as faint grey symbols.  The sample comprises galaxies with ages in the range $7.0\le\log(\textrm{Age/yr})\le9.6$, with the majority of galaxies having stellar metallicities of $0.0005\le Z_* \le 0.004$. }
    \label{fig:individualvssnr}
\end{figure}

\begin{figure*}
    \centering
    \includegraphics[width=1.0\linewidth]{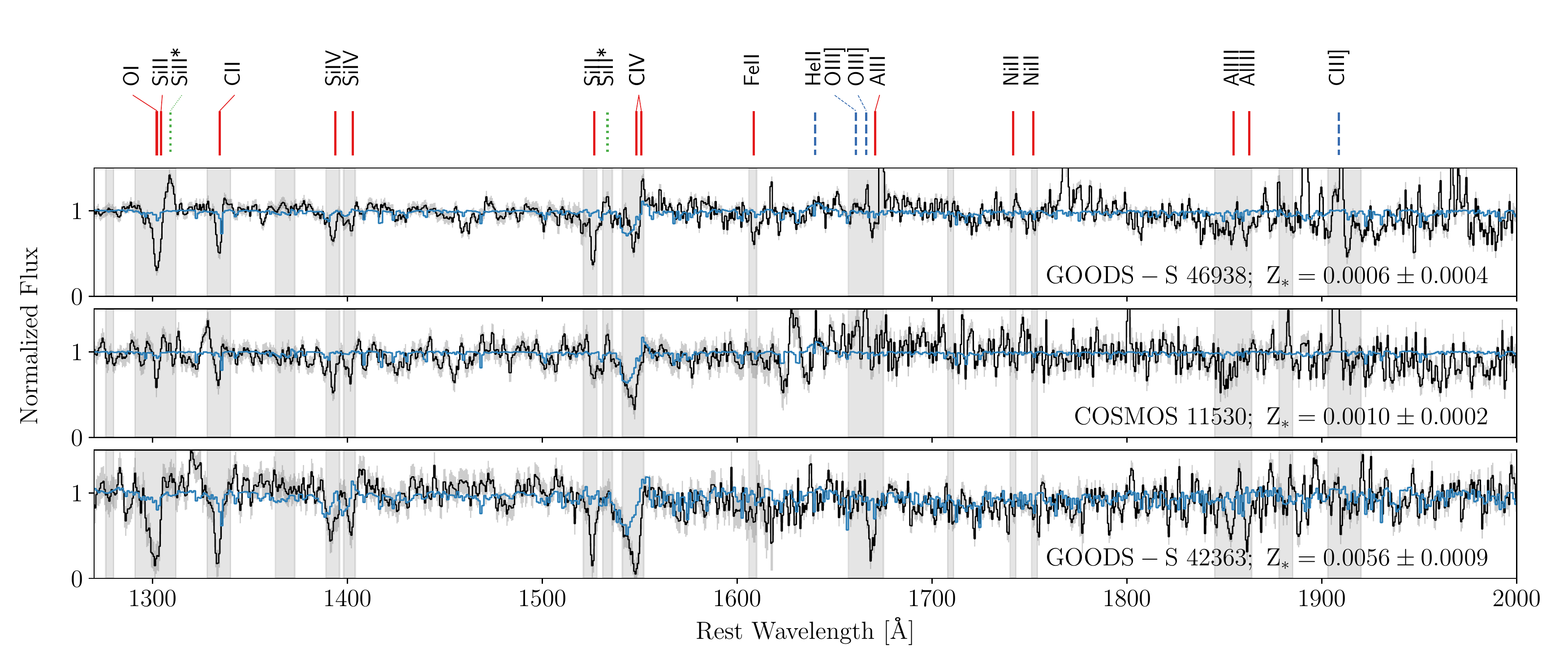}
    \caption{Example individual galaxy spectra (black) shown with their best-fit BPASS stellar population models (blue). These three spectra are examples of galaxies in our sample with low (top), close to median (middle), and high (bottom) stellar metallicities. The gray bars show regions of the spectra contaminated by interstellar absorption features that were excluded from fitting. Labels on the top of the figure indicate the locations of stellar absorption lines (solid red lines), nebular emission lines (dashed dark blue lines), and fine structure emission lines (dotted green lines). The $1\sigma$ error spectrum is depicted by the gray shaded region surrounding each spectrum.}
    \label{fig:examplefits}
\end{figure*}


\subsection{Testing models with additional SFHs}

We expanded the model grid used in \citet{Topping2020}, which included only stellar population models assuming a constant SFH. We repeated fitting the model grids to the \high and \low stacks using our updated models that assume different SFHs. For each SFH, we found that the results are consistent with those of \citet{Topping2020}. In particular, we fit models that assume a `delayed-$\tau$' SFH, with $\log(\tau/\textrm{year})=7, 8, 9, \textrm{and}\ 10$.  Figure~\ref{fig:bptallstellarfits} shows the best-fit age and stellar metallicity of the \textit{high} and \textit{low} stacks for each model grid.  For each SFH, we find that the \textit{high} stack has lower stellar metallicity, and a younger stellar age compared to the \textit{low} stack.  While this trend between the properties of the \textit{high} and \textit{low} stacks persists for each SFH we considered, the exact values of the stellar metallicity and age differ between the assumed models.  In particular, the delayed-$\tau$ model with $\log( \tau/\textrm{yr})=7$ has the youngest best-fit age and stellar metallicity, and both the age and stellar metallicity increase when assuming an increasing $\tau$.

For each of the assumed SFHs, we recovered the same qualitative trend found in \citet{Topping2020}, according to which the \high stack had a younger age and lower stellar metallicity relative to the \low stack.  This result assumes that both the \high and \low stacks are both described by a similar SFH, which is a reasonable assumption based on the SFHs determined for galaxies in these two stacks estimated from SED fitting. We also investigated if the stellar population models yielded any constraint on the form of the SFH for each stack. We tested this question by measuring the minimum $\chi^2$ value for the best-fit model of each SFH. For the \high and \low stacks, none of the SFHs were preferred over the others, suggesting that a given UV stellar spectrum is not unique to a particular SFH.  A KS-test revealed that the \textit{high} and \textit{low} stacks have $\tau$ distributions with a 45\% probability of being drawn from the same parent distribution.  Figure~\ref{fig:bestfitspectra} compares the best-fit spectrum of a constant SFH model to models with a delayed-$\tau$ SFH. The best-fit models for the \high stack are nearly identical, and at some wavelengths are different at the few percent level.  The \low stack models vary more, in particular for the $\log(\tau/\textrm{yr})=7$ model, which has some signatures of a young population not seen in the other best-fit models.  In particular, this model has slightly enhanced \CIV and \HeII emission compared to the other models.  However, the majority of the models are in agreement, with differences of only up to $\sim10\%$ in a few wavelength elements.  These tests have shown that we cannot determine which SFH best characterizes the rest-UV spectra. However, using models that assume different SFHs does not qualitatively affect our results.

\begin{figure*}
    \centering
    \includegraphics[width=1.0\linewidth]{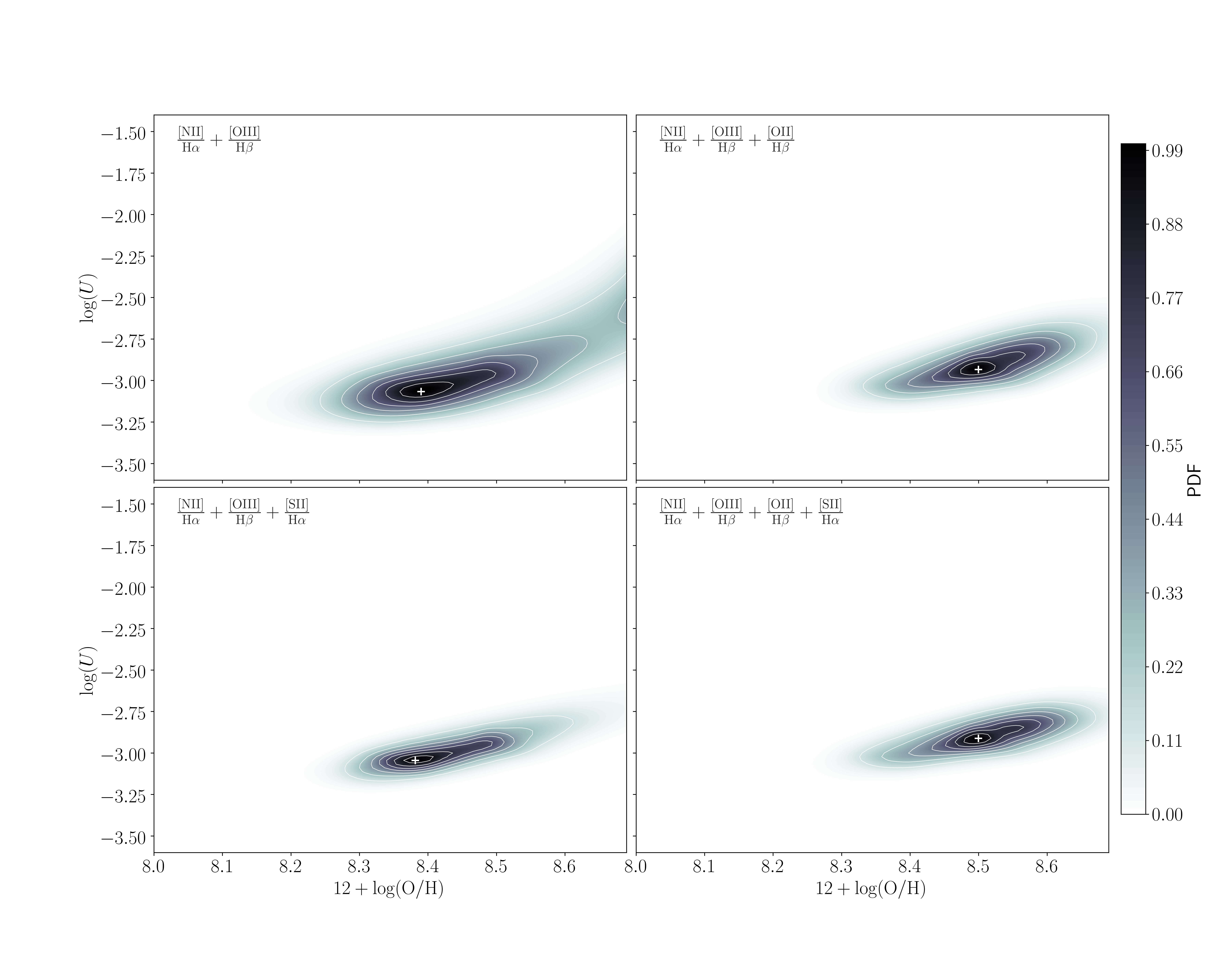}
    \caption{Probability density functions for inferring the ionization parameter ($\log(U)$) and nebular metallicity ($Z_{\textrm{neb}}$) when different sets of rest-optical emission lines are used. The text in the top left of each panel displays which lines correspond to each PDF.  All of the emission line fluxes are scaled to the observed H$\beta$ flux.  These panels demonstrate that the ionization parameter and nebular metallicity are better constrained when lines beyond [NII]$\lambda 6584$, [OIII]$\lambda 5007$, and H$\alpha$ are included in the fitting procedure.}
    \label{fig:nebularpdfs}
\end{figure*}

\subsection{The low SNR boundary to avoid biased results}

\begin{figure*}
    \centering
    \includegraphics[width=1.0\linewidth]{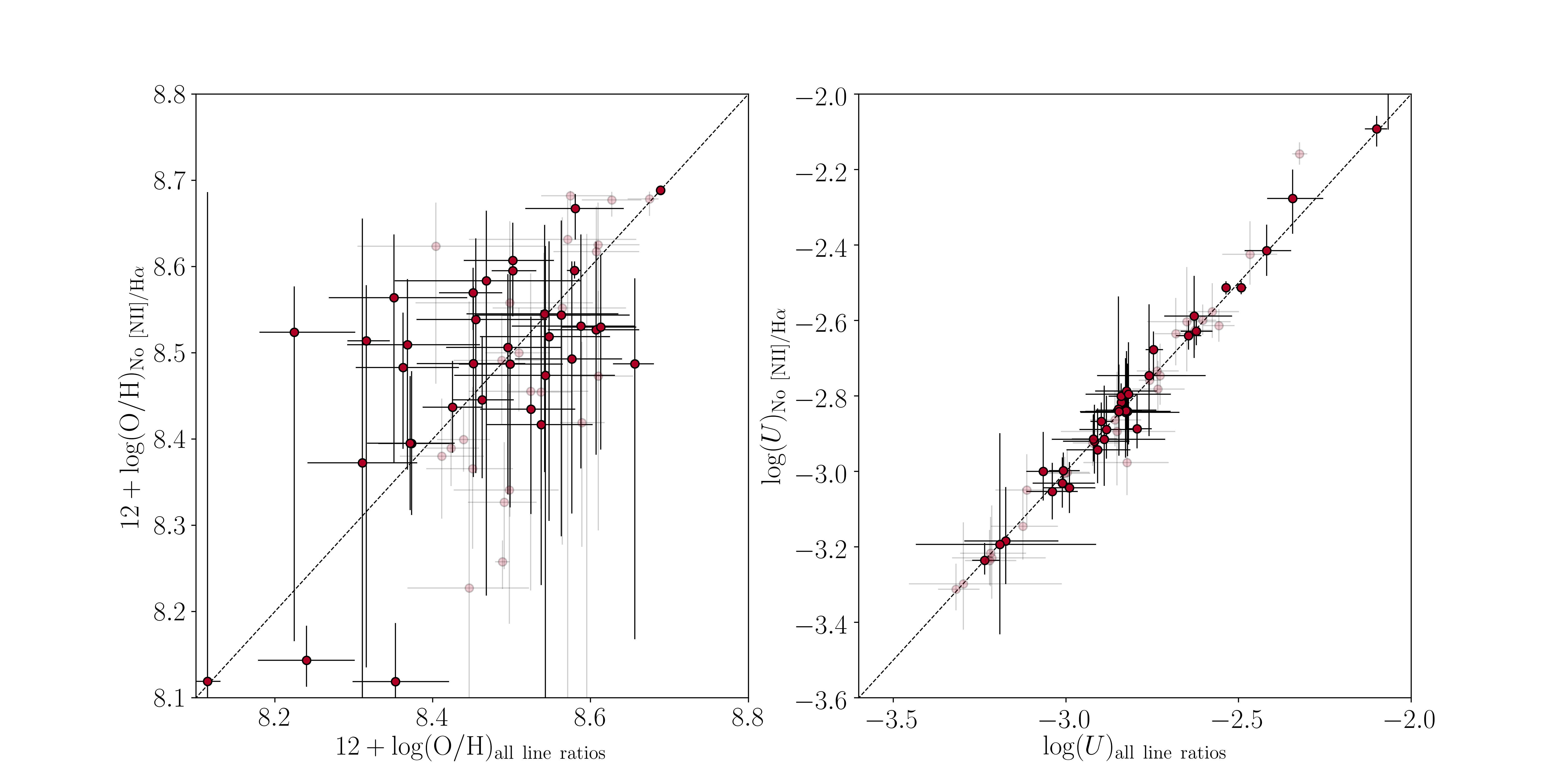}
    \caption{Comparison between inferred nebular parameters when [NII]$\lambda 6584/ \rm H\alpha$ is excluded from the fitting procedure. The dotted line displays the one-to-one relation in each panel.  Left: Best-fit $12+\log(\textrm{O/H})$ inferred without [NII]$\lambda 6584/ \rm H\alpha$ vs. $12+\log(\textrm{O/H})$ inferred with it included.  The majority of galaxies are consistent within their uncertainties using both methods.  Right: Same as the left panel except for $\log(U)$.  The values inferred with and without [NII]$\lambda 6584/\rm H\alpha$ agree remarkably well for nearly all galaxies.}
    \label{fig:Ninclude}
\end{figure*}

While the composite spectra provide useful constraints due to their high SNR, the SNR of individual galaxy spectra can be much lower. For example, the \textit{high} and \textit{low} stacks have $\rm SNR/pixel\sim 18$, while the individual spectra have a median $\rm SNR/pixel\sim 4.5$, where the sampling of our spectra are $\sim0.8\rm \angstrom /pixel$ in the rest frame. These values correspond to $\rm SNR\ per\ spectral\ resolution\ element \sim 25.5$ and $\rm SNR\ per\ spectral\ resolution\ element \sim 6.3$ for the composite and median individual spectra respectively.  We tested how the SNR of a spectrum affects the best-fit stellar properties by manually introducing noise to one of our composite spectra, refitting the models, and checking if biases arise as the spectrum drops in quality.  Figure~\ref{fig:stackscaledsnr} displays how the best-fit stellar metallicity and age change as a function of the amount of added noise.  For this composite, the best-fit stellar metallicity retains an unbiased estimate of the value obtained in the high-SNR limit down to a $\rm SNR\ per\ spectral\ resolution\ element \sim 5.6$ (SNR/pixel$\sim 4$). Below this value, the measurement of stellar metallicity is biased high. However, even above this limit, the stellar metallicity uncertainty increases with decreasing SNR.  The  best-fit age remains consistent throughout the range of SNR per spectral resolution element, yet as the SNR decreases below $\rm SNR\ per\ spectral\ resolution\ element \sim 5.6$, the uncertainty grows to $\ge2$ dex, leaving the age unconstrained.

This test showcases how biases in the best-fit stellar parameters may occur in lower SNR spectra. In order to further quantify this effect, we repeated the process used on the composite spectrum, except on BPASS models for which the `true' parameters are known.  We added noise selected from a normal distribution to the BPASS models at each wavelength element.  We repeated this for a combination of ages and stellar metallicities to determine if these biases exist throughout the range of parameter space our SFHs span.  Figure~\ref{fig:modelscaledsnr} shows how the best-fit age and stellar metallicity changes as noise is introduced into the model spectra.  At all stellar metallicities, a low SNR/resolution element introduces a positive bias in stellar metallicity leading to an overestimate of $Z_*$ relative to the true value.  At the lowest SNR, this bias can be up to $\sim150\%$. The best-fit stellar age also changes at low SNR/resolution element, however in contrast to the bias of the stellar metallicity, the trend of the bias depends on the `true' age.  At low SNR/resolution element, models with an old input age are biased younger, and models with a young age are biased toward older values.  While these biases exist at low SNR/resolution element for both age and stellar metallicity, the input values of $Z_*$ and age can be accurately determined for spectra with SNR/resolution element $\ge \sim5.6$.  Attempting to model the spectra with lower SNR leads to high uncertainties and systematic biases.

\section{Results}
\label{sec:results}

Based on our tests on the BPASS models, we can achieve accurate age (i.e., time since the onset of star formation) and stellar metallicity measurements for individual spectra with $\rm SNR\ per\ spectral\ resolution\ element \ge 5.6$. Figure~\ref{fig:individualvssnr} displays these values for all galaxies in the LRIS-BPT sample, highlighting those with sufficiently high SNR to yield unbiased $Z_*$ and age.  The stellar metallicity for the individual galaxies ranges between $0.001 \le Z_* \le 0.006$, consistent with the best-fit metallicities found for our composite spectra. Figure~\ref{fig:examplefits} shows some of our individual galaxy spectra and the best-fit BPASS model to each spectrum.  This figure shows the variety of stellar parameters recovered by our fitting by depicting a low, medium, and high stellar metallicity. Importantly, based on the stellar metallicity and age we are able to constrain the shape of the ionizing spectrum for each of these individual galaxies. However, for some galaxy spectra, the stellar metallicity probability distribution uniformly spanned the parameter space, and we could not assign a stellar metallicity in such cases.

\subsection{Ionized Gas Properties}

The ionizing spectrum emitted from the stellar population drives the production of the emergent rest-optical emission line ratios. Therefore, constraining the ionizing spectrum within galaxies is crucial in order to use photoionization models to extract physical properties from the observed nebular emission lines.  In particular, using Cloudy, we set the ionizing spectrum based on the appropriate best-fit BPASS stellar population model, and vary the nebular metallicity and ionization parameter, recording the emergent line fluxes.  We then compare the resulting catalog of nebular emission line flux ratios to those observed from an individual or composite spectrum. The inferred nebular metallicity and ionization parameter is set by which model best reproduces the observed emission lines defined by the minimum $\chi^2$ calculated for all observed line ratios simultaneously.  To understand the uncertainty in these quantities, we perturb the observed emission line flux ratios by their corresponding uncertainties, and recompute the best-fit nebular parameters.  \citet{Topping2020} used an approach that compared the locations of the models and observed galaxies on the [NII]$\lambda 6584$/H$\alpha$ vs. [OIII]$\lambda 5007$/H$\beta$ BPT diagram. In this analysis, we use a slightly different approach that simultaneously fits the [NII]$\lambda 6584$/H$\alpha$, and [OIII]$\lambda 5007$/H$\beta$ emission line ratios. Furthermore, we include additional strong line ratios, [SII]$\lambda \lambda 6717,6731$/H$\alpha$ and  [OII]$\lambda3727$/H$\beta$, in order to better constrain the nebular parameters. Because [OII]$\lambda3727$ and H$\beta$ are farther apart in wavelength compared to the other line ratios, we apply a dust correction to both lines.  We investigate the effect that including these additional observables has on the inferred nebular parameters. Figure~\ref{fig:nebularpdfs} shows an example of how the constraint on nebular metallicity and ionization parameter changes for different sets of nebular emission lines when our method is applied to the \textit{low} stack composite spectrum. In this example, the inferred nebular properties are consistent when considering different sets of lines, however the ionization parameter and nebular metallicity are better constrained when additional emission lines are included.

One assumption made in the photoionization modelling is the form of the N/O vs. O/H relation. The median nitrogen abundance of HII regions in the local universe has been measured to vary by $\sim 0.5$ dex for $8.2 \le 12+\log(\textrm{O/H}) \le 8.6$, with scatter of $\sim0.2$ dex at fixed O/H \citep{Pilyugin2012}.  This assumption strongly affects the output [NII]$\lambda 6584$/H$\alpha$ flux in our photoionization models.  These Nitrogen abundance variations can result in a disparity of the [NII]$\lambda 6584$/H$\alpha$ ratio $\sim 0.5$ dex, resulting in a biased inference of $Z_{\textrm{neb}}$ and $\log(U)$. Figure~\ref{fig:Ninclude} shows the effect of removing [NII]$\lambda 6584$/H$\alpha$ from our fitting procedure, eliminating the uncertainty surrounding the N/O relation.  Without [NII]$\lambda 6584$/H$\alpha$,  the inferred ionization parameters are well matched to those inferred when using [NII]$\lambda 6584$/H$\alpha$, with nearly all galaxies falling on the one-to-one relation.  In addition, there is a slight difference at low O/H when [NII]$\lambda 6584$/H$\alpha$ is not considered, however this difference is small compared to the uncertainties. Based on this result, we conclude that the inclusion of [NII]$\lambda 6584$/H$\alpha$ in our analysis does not significantly bias our inferred nebular parameters.

\begin{figure*}
    \centering
    \includegraphics[width=1.0\textwidth]{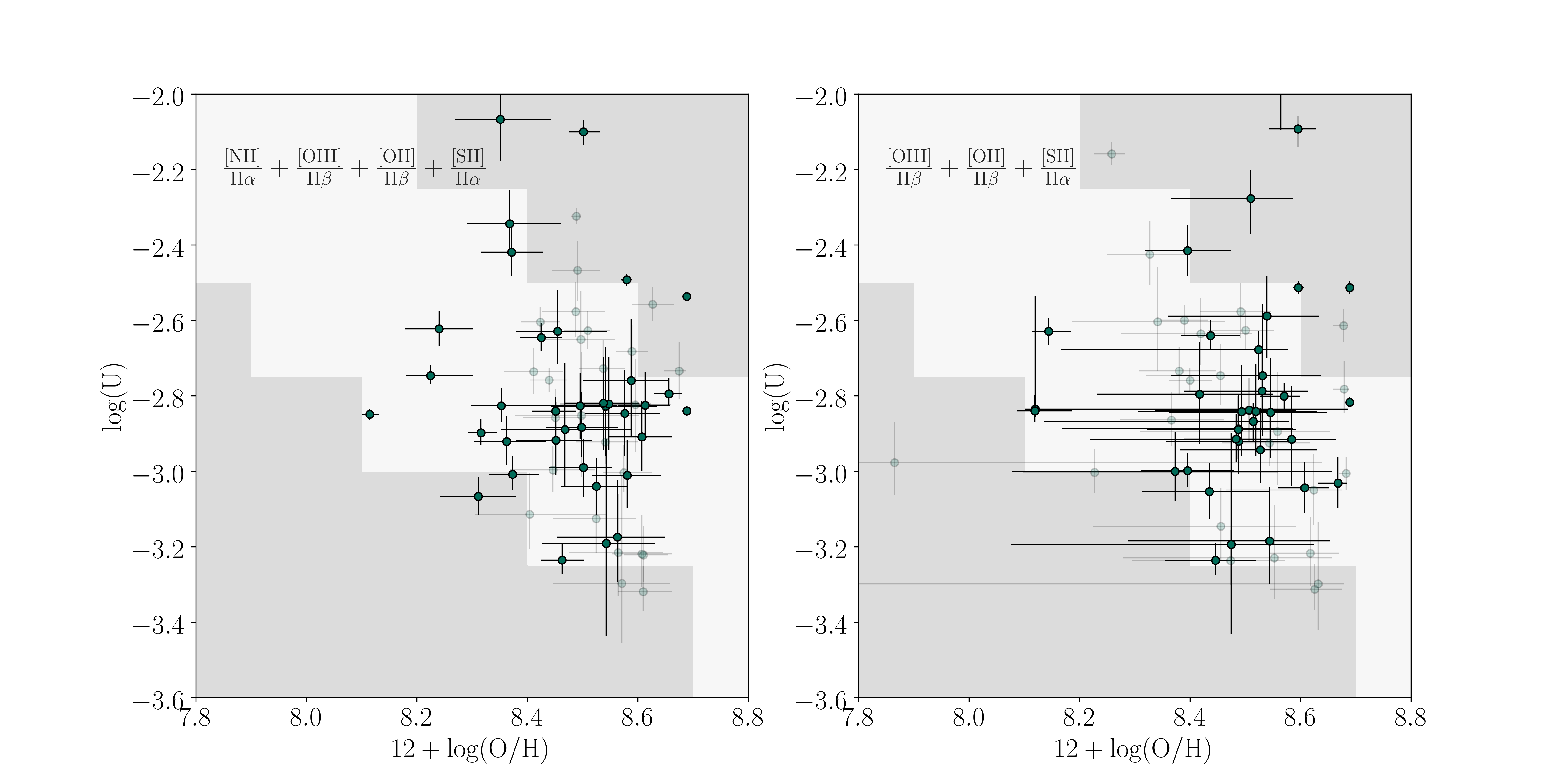}
    \caption{Inferred $\log(U)$ vs. $12+\log(\textrm{O/H})$ for each galaxy in the LRIS-BPT sample.  Galaxies with low SNR/resolution element ($\le5.6$) are displayed as the faint points.  The white region depicts the location of local HII regions in the parameter space as defined by \citet{PerezMontero2014}. Left: $\log(U)$ and $12+\log(\rm O/H)$ inferred using all strong rest-optical emission line ratios. The majority of galaxies in our sample lie within the region populated by local HII regions. Right: $\log(U)$ and $12+\log(\rm O/H)$ inferred when [NII]$\lambda 6584$/H$\alpha$ is excluded from our analysis.}
    \label{fig:nebparams}
\end{figure*}

\begin{figure}
    \centering
    \includegraphics[width=1.0\linewidth]{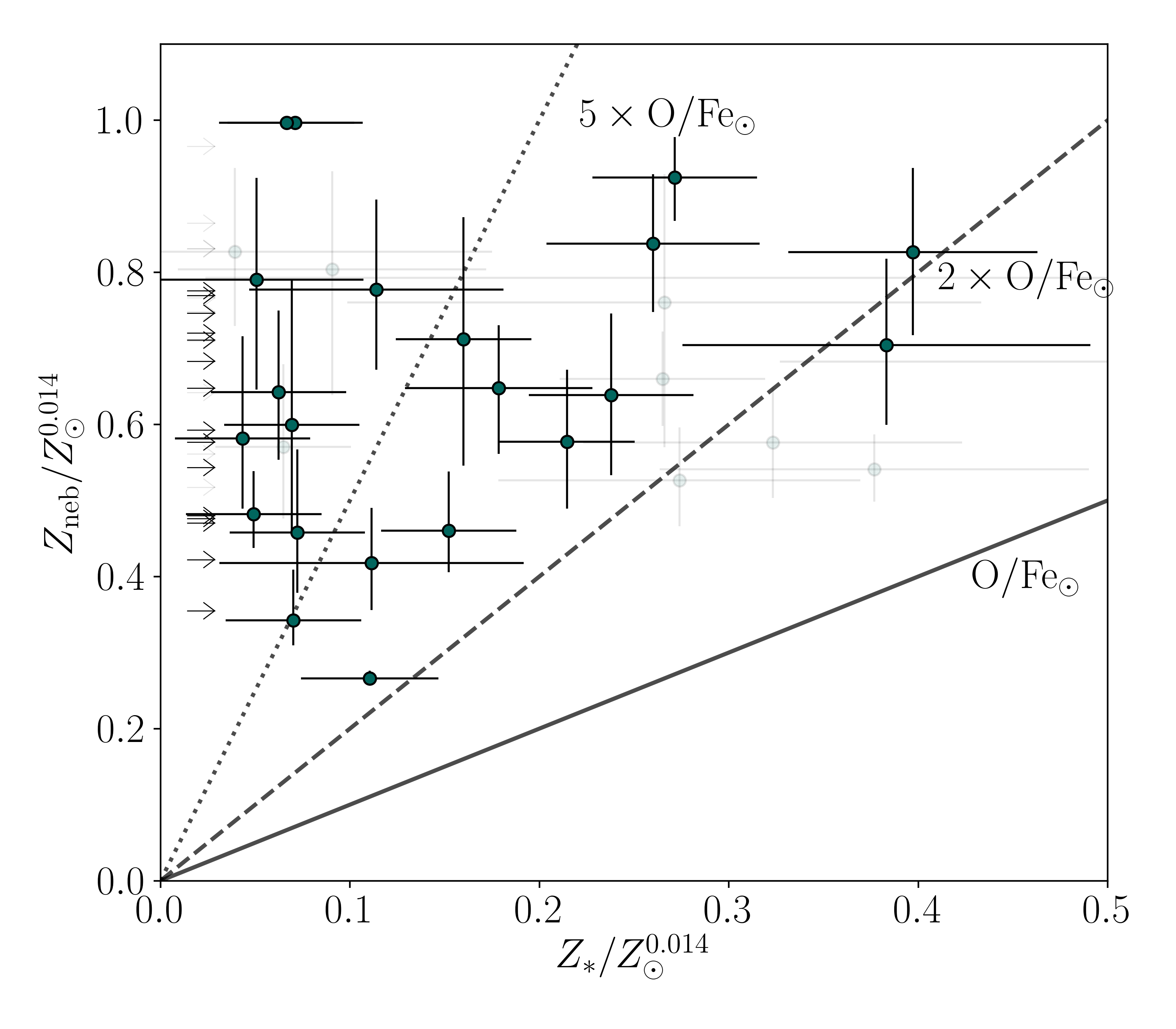}
    \caption{Nebular metallicity inferred from photoionization modelling plotted against stellar metallicity measured for each galaxy in our sample.  Lines of constant $\alpha$-enhancement (i.e., O/Fe) are displayed as solid, dashed, and dotted lines, respectively, for O/Fe$_{\odot}$, 2$\times$O/Fe$_{\odot}$, and 5$\times$O/Fe$_{\odot}$. All galaxies in our sample show evidence for $\alpha$-enhancement.  Additionally, some galaxies are in the regime above 5$\times$O/Fe$_{\odot}$, which has been suggested as the theoretical limit based on supernova yield models.  The galaxies for which the stellar metallicity could not be determined are displayed as lower limits.}
    \label{fig:alphaenh}
\end{figure}

Figure~\ref{fig:nebparams} shows $\log(U)$ against $12+\log(\textrm{O/H})$ for individual galaxies in the LRIS-BPT sample.  A majority of the galaxies in our high-SNR sample fall within the region populated by local HII regions shown by the white area in Figure~\ref{fig:nebparams} \citep{PerezMontero2014}, with median values of $12+\log(\textrm{O/H})=8.48\pm0.11$ and $\log(U)=-2.98\pm0.25$.  \citet{Sanders2020} found a median $\log(U) = -2.30\pm0.06$ for $z\sim2$ galaxies at $12+\log(\rm O/H)\sim8.0$, also consistent with the local U-O/H relation.  The measurements of $\log(U)$ and $12+\log(\textrm{O/H})$ for the galaxies in this high-SNR sample in combination with measurements of lower metallicity galaxies from \citet{Sanders2020} suggest that there may be a trend of decreasing ionization parameter with increasing nebular metallicity consistent with the trend seen in \citet{PerezMontero2014}. The offset of this sample from the $z\sim0$ excitation sequence in the BPT diagram is thus not driven by higher ionization parameters at fixed $\rm O/H$ \citep{Cullen2016, Kashino2017, Bian2020}. Furthermore, this result remains largely the same when considering $\log(U)$ and $12+\log(\textrm{O/H})$ inferred without [NII]$\lambda 6584/\rm H\alpha$. However, the different methods used to measure the oxygen abundances between our high-SNR sample, and those from \citet{PerezMontero2014}, which used the `direct' method may introduce systematics. In particular, \citet{Esteban2014} demonstrated that metallicities measured using the direct method are $\sim0.24$ lower than those which use nebular recombination lines on average.  However, metallicities measured from recombination lines are in better agreement with those inferred from photoionization modelling.  In addition the ionized gas properties presented by \citet{PerezMontero2014} were inferred using softer ionizing spectra compared to those used in this paper.  The result is a potential overestimate of the ionization parameter measured for the local systems compared to galaxies in our high-SNR sample.  While addressing this potential systematic could move our measurements away from the \citet{PerezMontero2014} relation, the difference in methods used to infer oxygen abundances would act in such a way to remedy the disparity.  Therefore, despite these systematics, we conclude that there is no strong offset in the U vs. O/H relation of galaxies in our sample compared to the local relation

\subsection{Combined Stellar and Nebular Properties}

To connect the stellar and nebular properties of the individual galaxies in our high-SNR sample, we combine the nebular O/H abundance inferred from photoionization modelling with the Fe/H measured from the BPASS model fitting to look at the $\alpha$-enhancement of individual galaxies.  Figure~\ref{fig:alphaenh} compares the nebular metallicity and the stellar metallicity for each galaxy in our sample.  Noticeably, all of the galaxies in our sample show evidence for $\alpha$-enhancement, falling significantly above the $\rm O/Fe_{\odot}$ line (solid black line). These values range from $\sim1.75\textrm{O/Fe}_{\odot}$ to $\ge 5\textrm{O/Fe}_{\odot}$. Investigating a small sample of $z\sim2-3.5$ galaxies with direct oxygen abundance measurements, \citet{Sanders2020} found similar results, where most of the galaxies analyzed show evidence for $\alpha$-enhancement, some having values $> 5\times\rm O/Fe_{\odot}$. A number of objects in our sample fall above the expected theoretical limit from pure Type II SNe of $\sim 5\times\rm O/Fe_{\odot}$, assuming a typical IMF \citep{Nomoto2006, Kobayashi2006}.  However, this limit depends on the details of the stellar population and expected Type II SNe yields. For example, the theoretical O/Fe limit increases when calculated assuming a top-heavy IMF.  Therefore, different assumptions of the IMF or supernova yields could remove the tension between the theoretical limit and some of our observed galaxies. Finally, the tension between our observed O/Fe and the theoretical limit may be relieved due to uncertainties in the stellar modelling of low-metallicity massive stars.  If, in fact, the low-metallicity stars produce harder ionizing radiation than what is included in the BPASS models, our galaxies would be best described by stellar metallicities that are higher than what we have found.  A higher best-fit stellar metallicity could bring the $\alpha$-enhancement into better agreement with the theoretical limits.

\begin{figure}
    \centering
    \includegraphics[width=1.0\linewidth]{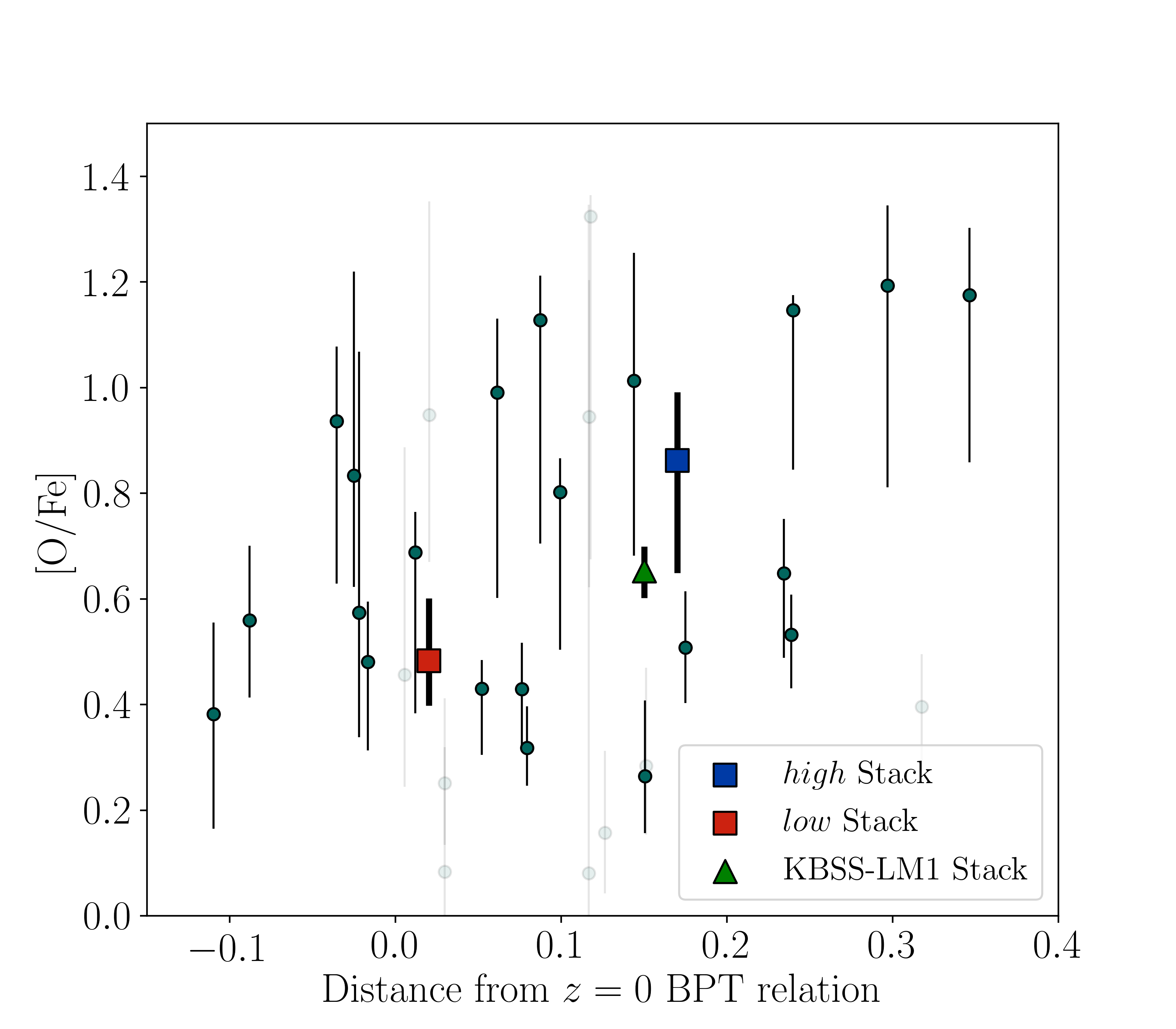}
    \caption{$\alpha$-enhancement calculated as a function of distance from the $z=0$ BPT star-forming sequence.  The galaxies that show the highest values of O/Fe are those that are the most offset from the local BPT sequence.  Measurements of the \high and \low stacks from \citet{Topping2020} are displayed as the blue and red squared respectively.  The KBSS-LM1 composite from \citet{Steidel2016} is shown as the green triangle.}
    \label{fig:alpha-bpt}
\end{figure}

We check how the $\alpha$-enhancement changes as a function of distance away from the local BPT sequence for our sample of high-redshift galaxies.  We calculate the distance away from the local sequence by first finding the line that is perpendicular to the local sequence from \citet{Kewley2013} defined as: 

\begin{equation}
\log(\rm [OIII]/H\beta) = \frac{0.61}{\log(\rm [NII]/H\alpha)+0.08}+1.1,
\end{equation}

\noindent and intersects the [NII]/H$\alpha$ and [OIII]/H$\beta$ value of our data point. The distance, D, is then:

\begin{equation}
    D^2 =\left(\rm N2-N2_{z=0}\right)^2 + \left(\rm O3-O3_{z=0}\right)^2,  
\end{equation}
where $\rm N2=\log(\rm [NII]\lambda 6584/H\alpha)$ and $\rm O3=\log([OIII]\lambda 5007/H\beta)$. We calculated uncertainties by perturbing each rest-UV spectrum and fitting the grid of BPASS stellar population models to measure the stellar metallicity.  The best-fit stellar population parameters define an ionizing spectrum which we then use as an input to Cloudy, from which, in turn, we infer the nebular oxygen abundance.  We define the $1\sigma$ uncertainty as the $16^{th}$ and $84^{th}$ percentile of the distribution of $\alpha$-enhancements resulting from repeating this process 1000 times. Figure~\ref{fig:alpha-bpt} displays how the $\alpha$-enhancement depends on our measured distance from the local BPT sequence. There appears to be a weak positive correlation between the distance away from the local BPT sequence and $\alpha$-enhancement.  While there is a large amount of scatter in this relation, we do see that the galaxies with the highest O/Fe typically lie farther from the local sequence.  We also compare the measurements of individual galaxies in our high-SNR sample with the \high and \low composite spectra from \citet{Topping2020}.  These two composite spectra follow the trend of higher O/Fe at a greater BPT distance, and both composite spectra are consistent with the results of our high-SNR sample.  While the data are not constraining enough to measure a functional form of a relation between $\alpha$-enhancement and BPT distance, they are consistent with the result that the offset of high-redshift galaxies relative to local galaxies on the BPT sequence is significantly driven by a harder ionizing spectrum due to $\alpha$-enhancement.

\section{Discussion}
Since the first evidence that suggested that high-redshift galaxies are offset on the BPT diagram, several hypotheses have been proposed to explain the underlying cause.  Among the proposed sources for this offset between local and high-redshift galaxies are harder ionizing spectra at fixed nebular metallicity, higher electron densities, contributions from AGNs and shocks at high redshift, and variations in gas-phase abundance patterns.  Recently, two prevailing theories suggest that the offset is primarily driven by higher ioniziation parameters at fixed gas-phase metallicity \citep{Kewley2015,Kashino2017,Cullen2018,Bian2018}, or that high-redshift galaxies exhibit a harder intrinsic ionizing spectrum at fixed nebular metallicity driven by $\alpha$-enhancement at high redshift \citep{Steidel2016, Sanders2020}.

To answer this question of the origin of the BPT offset, \citet{Sanders2020} used the `direct' method to estimate oxygen abundances for a sample of 18 high-redshift galaxies at low nebular metallicities. These authors then used photoionization models very similar to our own to fit for $\log(U)$ and $Z_*$ after fixing $Z_{\rm neb}$ to match the direct metallicity, finding that the average of the sample lies along $\log{U}$ vs. $12+\log(\rm O/H)$ of local HII regions \citep{PerezMontero2014}. This result suggests that the high ionization parameter measured in their sample is due to their low nebular metallicity, and that their sample has consistent ionization parameter with local HII regions that share the same O/H.  Furthermore, \citet{Shapley2019} demonstrated that high-redshift galaxies are also offset toward higher [SII]$\lambda \lambda 6717,6731/\textrm{H}\alpha$ and [OIII]$\lambda 5007$/H$\beta$ when using the appropriate comparison to local galaxies with low contribution from diffuse ionized gas (DIG) within the ISM. Using photoionization models from \citet{Sanders2016}, \citet{Shapley2019} concluded that both the offset on the [NII] and [SII] BPT diagrams is best explained by a harder ionizing spectrum at fixed nebular metallicity.

The results described in this paper suggest that $z\sim2$ galaxies do not have an elevated ionization parameter compared to local HII regions that share the same $12+\log(\textrm{O/H})$. Our analysis illustrates the importance of an independently constrained ionizing spectrum when trying to extract ionization parameter information from strong rest-optical line ratios. Without such a constraint, the degeneracy between ionization parameter and the intrinsic ionizing spectrum can bias inferences of the ionization parameter. It is important to note that the method used to infer oxygen abundances of our sample is different from the method of our local HII region comparison sample \citep{PerezMontero2014}, which could introduce systematics in the comparison. The offset between these two methods results in $\sim0.24$ dex lower oxygen abundance when using the direct method relative to the nebular recombination lines (see, e.g., \citet{Sanders2020} for a more detailed discussion). However, the modelling from \citet{PerezMontero2014} assumed a softer ionizing spectrum at fixed oxygen abundance compared relative to the BPASS models used for our analysis.  Using a softer ionizing spectrum results in an ionization parameter that is systematically higher for a fixed set of nebular emission line ratios \citep{Sanders2016}. Therefore, correcting for these biases affecting the inferred ionization parameter and oxygen abundance would shift our inferred values along the U vs. O/H relation described by \citet{PerezMontero2014}, and would not lead to a significant difference between the nebular parameters of local HII regions and the $z\sim2$ galaxies in our sample. 

Finally, we address the possibility of differing abundance scales between our stellar and nebular fitting procedures using a few different methods.  We remeasured the best-fit stellar metallicities using Starburst99 stellar population synthesis models \citep{Leitherer2014} and compared the results to those found using BPASS models.  In the range of stellar metallicities spanned by the majority of galaxies in our sample, ($0.001\lesssim Z_* \lesssim 0.004$) the Starburst99- and BPASS-based stellar metallicities agree within the uncertainties.  However, for some galaxies the stellar metallicity found by BPASS is $<0.001$, the minimum metallicity in the Starburst99 grid.  In addition, both BPASS and Starburst99 models have only been computed using solar abundance patterns.  For this analysis, altering the abundance patterns within the models is unlikely to significantly affect our results, as the differences between models of varied stellar metallicities is primarily driven by the Fe abundance \citep{Steidel2016}, while the other elements have subdominant effects.  To address the gas-phase abundance scale, the ideal case would be to utilize direct-method metallicities.  However, making the necessary measurements to calculate direct metallicities is challenging \citep{Sanders2020}, and requires a data set that does not exist for a large sample and in the metallicity range spanned by our galaxies.  Instead, we compared the inferred nebular properties presented above using photoionization modelling to properties inferred using an empirical method described by \citet{Sanders2020} using the calibrations from \citet{Bian2018}.  We find that the empirical measurements of $12+\log(\rm O/H)$ are on average $0.12$ dex lower than those inferred from photoionization modelling. Despite this difference, when using oxygen abundances inferred from empirical calibrations we find that all of our galaxies show evidence for $\alpha$-enhancement. One other consideration is the possibility of rapid oxygen enrichment of the HII regions resulting in a different nebular abundance relative to that of the massive stars.  However, analysis of the Orion star-forming region finds the gas-phase oxygen abundance is in reasonable agreement with the oxygen abundance massive stars \citep{Simondiaz2010, Simondiaz2011}. Based on our stellar and nebular results, we find that the offset on the BPT diagram is primarily due to a harder ionizing spectrum resulting from sub-solar stellar Fe abundances relative to local galaxies.

\section{Summary \& Conclusions}
\label{sec:conclusions}

We used the combination of rest-UV and rest-optical spectra for a sample of 62 galaxies to investigate the physical conditions within galaxies at $z\sim2.3$.  We expanded upon the results of \citet{Topping2020} in which we constructed composite spectra based on location in the [NII]$\lambda 6584$ BPT diagram, and found that galaxies offset from the local sequence typically had younger ages, lower stellar metallicities, higher ionization parameters, and were more $\alpha$-enhanced.  We expanded the fitting analysis to include additional SFHs and rest-optical emission line fluxes.  In addition, we quantitatively determined the rest-UV continuum SNR limit above which we can constrain the stellar metallicity and age of individual galaxies in an unbiased manner.  We summarize our main results and conclusions below.

(i) We constructed additional BPASS stellar population models for a variety of SFHs.  We repeated the fitting analysis of the two stacked spectra defined by \citet{Topping2020} and found that for each SFH, the stack composed of galaxies offset from the local sequence on the BPT diagram had a younger age and lower stellar metallicity.  Additionally, when fitting across all SFHs for a single stack, we do not find any preference for one SFH over another.  Therefore, we cannot determine which SFH best characterizes the rest-UV spectra.  

(ii) We tested the SNR down to which individual galaxy spectra are suitable to be fit using this type of analysis.  Based on the test of perturbing model spectra with known stellar parameters with increasing amounts of noise, we find that the best-fit stellar metallicity is biased high when the spectrum reaches a SNR/resolution element$ < 5.6$, and the best-fit age is biased toward the middle of the grid ($\log(\rm Age/yr) \sim8.5$), with an uncertainty that fills the parameter space. The best-fit age and stellar metallicity remain consistent with the true value for spectra with $\rm SNR\ per\ spectral\ resolution\ element > 5.6$. Therefore, rest-UV spectral fitting should not be attempted on spectra for which the SNR/resolution element is less than $5.6$ in order to avoid biased results.

(iii)  Based on the SNR requirements described above, we found that 30 galaxies in our sample satisfied the criteria to be fit on an individual basis.  We find that galaxies in this high-SNR sample have ages that have a wide ranges of ages spanning $\log(\rm Age/yr)\sim 7.0-9.6$, and that most galaxies have stellar metallicities in the range $\sim0.001 < Z_* < 0.004$.

(iv) We examined how different rest-optical emission lines affect the inferred ionization parameter and nebular metallicity.  Previously, we inferred nebular parameters by comparing observed [NII]$\lambda 6584$/H$\alpha$ and [OIII]$\lambda 5007$/H$\beta$ with a suite of photoionization models. In this analysis, we tested how adding [SII]$\lambda \lambda 6717,6731$/H$\alpha$ and [OII]$\lambda3727$/H$\beta$ to the fitting procedure affects the resultant parameters.  In general, adding the additional lines yields results with smaller uncertainties.  Additionally, because one assumption we made is the form of the N/O vs. O/H relation as an input to our models, we tested fitting the rest-optical lines that are not affected by this assumption, namely [OIII]$\lambda 5007$/H$\beta$, [OII]$\lambda3727$/H$\beta$, and [SII]$\lambda \lambda 6717,6731$/H$\alpha$.  We find that when [NII]$\rm \lambda 6584/H\alpha$ is excluded from fitting, the nebular metallicity and ionization parameter remain consistent with the values inferred when [NII]$\rm \lambda 6584/H\alpha$ is included.

(v) With the constrained ionizing spectrum for each individual galaxy, we used photoionization models to infer ionization parameters and nebular metallicities for each galaxy in our sample, and find that the inferred ionization parameters ($\log(U)_{\textrm{med}}=-2.98\pm0.25$) are consistent with those measured in local HII regions that share the same oxygen abundance ($12+\log(\textrm{O/H})_{\textrm{med}}=8.48\pm0.11$). This result suggests that the offset of high-redshift galaxies on the BPT diagram relative to local galaxies is not due to elevated ionization parameters at fixed $\rm O/H$.

(vi) Combining the best-fit stellar metallicities from fitting BPASS model spectra to the nebular metallicities inferred from photoionization modelling we find that all of our individual galaxies are $\alpha$-enhanced compared to local galaxies. Furthermore, we find that high-redshift galaxies that show the highest levels of $\alpha$-enhancement are typically more offset from the local BPT star-forming sequence. The O/Fe values span the range from $\sim1.75\times$O/Fe$_{\odot}$ to above the theoretical limit \citep[$\sim5\times$O/Fe$_{\odot}$][]{Nomoto2006}.  This limit could be affected by details of the IMF, including the high-mass slope and upper mass cutoff. Furthermore, the stellar metallicities may change as stellar modelling of the most metal-poor massive stars are better understood.  In particular, if the lowest $Z_*$ stars actually produce harder ionizing spectra compared to current models, we would infer a higher stellar metallicity for our rest-UV spectra. In combination with (v), this result suggests that the offset of high-redshift galaxies on the BPT diagram relative to the local sequence is likely due to a harder ionizing spectrum at fixed nebular oxygen abundance resulting from elevated O/Fe in high-redshift galaxies.

A combined understanding of the intrinsic ionizing spectrum that excites HII regions and the physical properties of the ISM is a crucial step toward a complete model of high-redshift galaxy evolution.  In order to fully understand high-redshift galaxies we must explore how their properties differ from local galaxies, but also how the population of high-redshift galaxies varies within itself.  Ultimately, detailed modelling of large numbers of individual galaxies will be required to expand our understanding of galaxies beyond the level of a sample average, but our work here begins to illustrate the range and interplay of nebular and stellar properties observed during the epoch of peak star formation in the universe.  Constraining the ionizing spectrum is crucial in order to use photoionization models to infer nebular properties, as without this constraint solutions are highly degenerate. Rest-UV spectroscopy is the ideal tool to gain insight into the massive star population, and therefore the ionizing spectrum, for individual galaxies at high redshift. This type of analysis is key in order to compare the internal properties of high-redshift galaxies to those of local HII regions and galaxies.

\section*{Acknowledgements}
 We thank the anonymous referee for their constructive comments. We acknowledge support from NSF AAG grants AST1312780, 1312547, 1312764, and 1313171, grant AR13907 from the Space Telescope Science Institute, and grant NNX16AF54G from the NASA ADAP program. We also acknowledge a NASA contract supporting the ``WFIRST Extragalactic Potential Observations (EXPO) Science Investigation Team'' (15-WFIRST15-0004), administered by GSFC. This work made use of v2.2.1 of the Binary Population and Spectral Synthesis (BPASS) models as described in Eldridge, Stanway et al. (2017) and Stanway \& Eldridge et al. (2018). We wish to extend special thanks to those of Hawaiian ancestry on whose sacred mountain we are privileged to be guests. Without their generous hospitality, most of the observations presented herein would not have been possible.
 
 \section*{Data Availability}
 The data underlying this article will be shared on reasonable request to the corresponding author.

\bibliographystyle{mnras}
\bibliography{mosdef_lris-modelling}

\begin{thebibliography}{}
\makeatletter
\relax
\def\mn@urlcharsother{\let\do\@makeother \do\$\do\&\do\#\do\^\do\_\do\%\do\~}
\def\mn@doi{\begingroup\mn@urlcharsother \@ifnextchar [ {\mn@doi@}
  {\mn@doi@[]}}
\def\mn@doi@[#1]#2{\def\@tempa{#1}\ifx\@tempa\@empty \href
  {http://dx.doi.org/#2} {doi:#2}\else \href {http://dx.doi.org/#2} {#1}\fi
  \endgroup}
\def\mn@eprint#1#2{\mn@eprint@#1:#2::\@nil}
\def\mn@eprint@arXiv#1{\href {http://arxiv.org/abs/#1} {{\tt arXiv:#1}}}
\def\mn@eprint@dblp#1{\href {http://dblp.uni-trier.de/rec/bibtex/#1.xml}
  {dblp:#1}}
\def\mn@eprint@#1:#2:#3:#4\@nil{\def\@tempa {#1}\def\@tempb {#2}\def\@tempc
  {#3}\ifx \@tempc \@empty \let \@tempc \@tempb \let \@tempb \@tempa \fi \ifx
  \@tempb \@empty \def\@tempb {arXiv}\fi \@ifundefined
  {mn@eprint@\@tempb}{\@tempb:\@tempc}{\expandafter \expandafter \csname
  mn@eprint@\@tempb\endcsname \expandafter{\@tempc}}}

\bibitem[\protect\citeauthoryear{{Abazajian} et~al.,}{{Abazajian}
  et~al.}{2009}]{Abazajian2009}
{Abazajian} K.~N.,  et~al., 2009, \mn@doi [\apjs]
  {10.1088/0067-0049/182/2/543}, \href
  {https://ui.adsabs.harvard.edu/abs/2009ApJS..182..543A} {182, 543}

\bibitem[\protect\citeauthoryear{{Asplund}, {Grevesse}, {Sauval}  \&
  {Scott}}{{Asplund} et~al.}{2009}]{Asplund2009}
{Asplund} M.,  {Grevesse} N.,  {Sauval} A.~J.,   {Scott} P.,  2009, \mn@doi
  [\araa] {10.1146/annurev.astro.46.060407.145222}, \href
  {https://ui.adsabs.harvard.edu/abs/2009ARA&A..47..481A} {47, 481}

\bibitem[\protect\citeauthoryear{{Baldwin}, {Phillips}  \&
  {Terlevich}}{{Baldwin} et~al.}{1981}]{Baldwin1981}
{Baldwin} J.~A.,  {Phillips} M.~M.,   {Terlevich} R.,  1981, \mn@doi [\pasp]
  {10.1086/130766}, \href
  {https://ui.adsabs.harvard.edu/abs/1981PASP...93....5B} {93, 5}

\bibitem[\protect\citeauthoryear{{Bian}, {Kewley}  \& {Dopita}}{{Bian}
  et~al.}{2018}]{Bian2018}
{Bian} F.,  {Kewley} L.~J.,   {Dopita} M.~A.,  2018, \mn@doi [\apj]
  {10.3847/1538-4357/aabd74}, \href
  {https://ui.adsabs.harvard.edu/abs/2018ApJ...859..175B} {859, 175}

\bibitem[\protect\citeauthoryear{{Bian}, {Kewley}, {Groves}  \&
  {Dopita}}{{Bian} et~al.}{2020}]{Bian2020}
{Bian} F.,  {Kewley} L.~J.,  {Groves} B.,   {Dopita} M.~A.,  2020, \mn@doi
  [\mnras] {10.1093/mnras/staa259}, \href
  {https://ui.adsabs.harvard.edu/abs/2020MNRAS.493..580B} {493, 580}

\bibitem[\protect\citeauthoryear{{Chabrier}}{{Chabrier}}{2003}]{Chabrier2003}
{Chabrier} G.,  2003, \mn@doi [\pasp] {10.1086/376392}, \href
  {https://ui.adsabs.harvard.edu/abs/2003PASP..115..763C} {115, 763}

\bibitem[\protect\citeauthoryear{{Chisholm}, {Rigby}, {Bayliss}, {Berg},
  {Dahle}, {Gladders}  \& {Sharon}}{{Chisholm} et~al.}{2019}]{Chisholm2019}
{Chisholm} J.,  {Rigby} J.~R.,  {Bayliss} M.,  {Berg} D.~A.,  {Dahle} H.,
  {Gladders} M.,   {Sharon} K.,  2019, \mn@doi [\apj]
  {10.3847/1538-4357/ab3104}, \href
  {https://ui.adsabs.harvard.edu/abs/2019ApJ...882..182C} {882, 182}

\bibitem[\protect\citeauthoryear{{Crowther}, {Prinja}, {Pettini}  \&
  {Steidel}}{{Crowther} et~al.}{2006}]{Crowther2006}
{Crowther} P.~A.,  {Prinja} R.~K.,  {Pettini} M.,   {Steidel} C.~C.,  2006,
  \mn@doi [\mnras] {10.1111/j.1365-2966.2006.10164.x}, \href
  {https://ui.adsabs.harvard.edu/abs/2006MNRAS.368..895C} {368, 895}

\bibitem[\protect\citeauthoryear{{Cullen}, {Cirasuolo}, {Kewley}, {McLure},
  {Dunlop}  \& {Bowler}}{{Cullen} et~al.}{2016}]{Cullen2016}
{Cullen} F.,  {Cirasuolo} M.,  {Kewley} L.~J.,  {McLure} R.~J.,  {Dunlop}
  J.~S.,   {Bowler} R.~A.~A.,  2016, \mn@doi [\mnras] {10.1093/mnras/stw1181},
  \href {https://ui.adsabs.harvard.edu/abs/2016MNRAS.460.3002C} {460, 3002}

\bibitem[\protect\citeauthoryear{{Cullen} et~al.,}{{Cullen}
  et~al.}{2018}]{Cullen2018}
{Cullen} F.,  et~al., 2018, \mn@doi [\mnras] {10.1093/mnras/sty469}, \href
  {https://ui.adsabs.harvard.edu/abs/2018MNRAS.476.3218C} {476, 3218}

\bibitem[\protect\citeauthoryear{{Cullen} et~al.,}{{Cullen}
  et~al.}{2019}]{Cullen2019}
{Cullen} F.,  et~al., 2019, \mn@doi [\mnras] {10.1093/mnras/stz1402}, \href
  {https://ui.adsabs.harvard.edu/abs/2019MNRAS.487.2038C} {487, 2038}

\bibitem[\protect\citeauthoryear{{Eldridge}, {Stanway}, {Xiao}, {McClelland },
  {Taylor}, {Ng}, {Greis}  \& {Bray}}{{Eldridge} et~al.}{2017}]{Eldridge2017}
{Eldridge} J.~J.,  {Stanway} E.~R.,  {Xiao} L.,  {McClelland } L.~A.~S.,
  {Taylor} G.,  {Ng} M.,  {Greis} S.~M.~L.,   {Bray} J.~C.,  2017, \mn@doi
  [\pasa] {10.1017/pasa.2017.51}, \href
  {https://ui.adsabs.harvard.edu/abs/2017PASA...34...58E} {34, e058}

\bibitem[\protect\citeauthoryear{{Erb}, {Shapley}, {Pettini}, {Steidel},
  {Reddy}  \& {Adelberger}}{{Erb} et~al.}{2006}]{Erb2006}
{Erb} D.~K.,  {Shapley} A.~E.,  {Pettini} M.,  {Steidel} C.~C.,  {Reddy} N.~A.,
    {Adelberger} K.~L.,  2006, \mn@doi [\apj] {10.1086/503623}, \href
  {https://ui.adsabs.harvard.edu/abs/2006ApJ...644..813E} {644, 813}

\bibitem[\protect\citeauthoryear{{Esteban}, {Garc{\'\i}a-Rojas}, {Carigi},
  {Peimbert}, {Bresolin}, {L{\'o}pez-S{\'a}nchez}  \& {Mesa-Delgado}}{{Esteban}
  et~al.}{2014}]{Esteban2014}
{Esteban} C.,  {Garc{\'\i}a-Rojas} J.,  {Carigi} L.,  {Peimbert} M.,
  {Bresolin} F.,  {L{\'o}pez-S{\'a}nchez} A.~R.,   {Mesa-Delgado} A.,  2014,
  \mn@doi [\mnras] {10.1093/mnras/stu1177}, \href
  {https://ui.adsabs.harvard.edu/abs/2014MNRAS.443..624E} {443, 624}

\bibitem[\protect\citeauthoryear{{Ferland} et~al.,}{{Ferland}
  et~al.}{2017}]{Ferland2017}
{Ferland} G.~J.,  et~al., 2017, Revista Mexicana de Astronomia y Astrofisica,
  \href {https://ui.adsabs.harvard.edu/abs/2017RMxAA..53..385F} {53, 385}

\bibitem[\protect\citeauthoryear{{Grogin} et~al.,}{{Grogin}
  et~al.}{2011}]{Grogin2011}
{Grogin} N.~A.,  et~al., 2011, \mn@doi [\apjs] {10.1088/0067-0049/197/2/35},
  \href {https://ui.adsabs.harvard.edu/abs/2011ApJS..197...35G} {197, 35}

\bibitem[\protect\citeauthoryear{{Halliday} et~al.,}{{Halliday}
  et~al.}{2008}]{Halliday2008}
{Halliday} C.,  et~al., 2008, \mn@doi [\aap] {10.1051/0004-6361:20078673},
  \href {https://ui.adsabs.harvard.edu/abs/2008A&A...479..417H} {479, 417}

\bibitem[\protect\citeauthoryear{{Kashino} et~al.,}{{Kashino}
  et~al.}{2017}]{Kashino2017}
{Kashino} D.,  et~al., 2017, \mn@doi [\apj] {10.3847/1538-4357/835/1/88}, \href
  {https://ui.adsabs.harvard.edu/abs/2017ApJ...835...88K} {835, 88}

\bibitem[\protect\citeauthoryear{{Kauffmann} et~al.,}{{Kauffmann}
  et~al.}{2003}]{Kauffmann2003}
{Kauffmann} G.,  et~al., 2003, \mn@doi [\mnras]
  {10.1111/j.1365-2966.2003.07154.x}, \href
  {https://ui.adsabs.harvard.edu/abs/2003MNRAS.346.1055K} {346, 1055}

\bibitem[\protect\citeauthoryear{{Kewley}, {Dopita}, {Sutherland}, {Heisler}
  \& {Trevena}}{{Kewley} et~al.}{2001}]{Kewley2001}
{Kewley} L.~J.,  {Dopita} M.~A.,  {Sutherland} R.~S.,  {Heisler} C.~A.,
  {Trevena} J.,  2001, \mn@doi [\apj] {10.1086/321545}, \href
  {https://ui.adsabs.harvard.edu/abs/2001ApJ...556..121K} {556, 121}

\bibitem[\protect\citeauthoryear{{Kewley}, {Dopita}, {Leitherer}, {Dav{\'e}},
  {Yuan}, {Allen}, {Groves}  \& {Sutherland}}{{Kewley}
  et~al.}{2013}]{Kewley2013}
{Kewley} L.~J.,  {Dopita} M.~A.,  {Leitherer} C.,  {Dav{\'e}} R.,  {Yuan} T.,
  {Allen} M.,  {Groves} B.,   {Sutherland} R.,  2013, \mn@doi [\apj]
  {10.1088/0004-637X/774/2/100}, \href
  {https://ui.adsabs.harvard.edu/abs/2013ApJ...774..100K} {774, 100}

\bibitem[\protect\citeauthoryear{{Kewley}, {Zahid}, {Geller}, {Dopita}, {Hwang}
   \& {Fabricant}}{{Kewley} et~al.}{2015}]{Kewley2015}
{Kewley} L.~J.,  {Zahid} H.~J.,  {Geller} M.~J.,  {Dopita} M.~A.,  {Hwang}
  H.~S.,   {Fabricant} D.,  2015, \mn@doi [\apjl]
  {10.1088/2041-8205/812/2/L20}, \href
  {https://ui.adsabs.harvard.edu/abs/2015ApJ...812L..20K} {812, L20}

\bibitem[\protect\citeauthoryear{{Kobayashi}, {Umeda}, {Nomoto}, {Tominaga}  \&
  {Ohkubo}}{{Kobayashi} et~al.}{2006}]{Kobayashi2006}
{Kobayashi} C.,  {Umeda} H.,  {Nomoto} K.,  {Tominaga} N.,   {Ohkubo} T.,
  2006, \mn@doi [\apj] {10.1086/508914}, \href
  {https://ui.adsabs.harvard.edu/abs/2006ApJ...653.1145K} {653, 1145}

\bibitem[\protect\citeauthoryear{{Kriek} et~al.,}{{Kriek}
  et~al.}{2015}]{Kriek2015}
{Kriek} M.,  et~al., 2015, \mn@doi [\apjs] {10.1088/0067-0049/218/2/15}, \href
  {https://ui.adsabs.harvard.edu/abs/2015ApJS..218...15K} {218, 15}

\bibitem[\protect\citeauthoryear{{Leitherer}, {Le{\~a}o}, {Heckman}, {Lennon},
  {Pettini}  \& {Robert}}{{Leitherer} et~al.}{2001}]{Leitherer2001}
{Leitherer} C.,  {Le{\~a}o} J. R.~S.,  {Heckman} T.~M.,  {Lennon} D.~J.,
  {Pettini} M.,   {Robert} C.,  2001, \mn@doi [\apj] {10.1086/319814}, \href
  {https://ui.adsabs.harvard.edu/abs/2001ApJ...550..724L} {550, 724}

\bibitem[\protect\citeauthoryear{{Leitherer}, {Ekstr{\"o}m}, {Meynet},
  {Schaerer}, {Agienko}  \& {Levesque}}{{Leitherer}
  et~al.}{2014}]{Leitherer2014}
{Leitherer} C.,  {Ekstr{\"o}m} S.,  {Meynet} G.,  {Schaerer} D.,  {Agienko}
  K.~B.,   {Levesque} E.~M.,  2014, \mn@doi [\apjs]
  {10.1088/0067-0049/212/1/14}, \href
  {https://ui.adsabs.harvard.edu/abs/2014ApJS..212...14L} {212, 14}

\bibitem[\protect\citeauthoryear{{Liu}, {Shapley}, {Coil}, {Brinchmann}  \&
  {Ma}}{{Liu} et~al.}{2008}]{Liu2008}
{Liu} X.,  {Shapley} A.~E.,  {Coil} A.~L.,  {Brinchmann} J.,   {Ma} C.-P.,
  2008, \mn@doi [\apj] {10.1086/529030}, \href
  {https://ui.adsabs.harvard.edu/abs/2008ApJ...678..758L} {678, 758}

\bibitem[\protect\citeauthoryear{{Masters} et~al.,}{{Masters}
  et~al.}{2014}]{Masters2014}
{Masters} D.,  et~al., 2014, \mn@doi [\apj] {10.1088/0004-637X/785/2/153},
  \href {https://ui.adsabs.harvard.edu/abs/2014ApJ...785..153M} {785, 153}

\bibitem[\protect\citeauthoryear{{Matthee} \& {Schaye}}{{Matthee} \&
  {Schaye}}{2018}]{Matthee2018}
{Matthee} J.,  {Schaye} J.,  2018, \mn@doi [\mnras] {10.1093/mnrasl/sly093},
  \href {https://ui.adsabs.harvard.edu/abs/2018MNRAS.479L..34M} {479, L34}

\bibitem[\protect\citeauthoryear{{McLean} et~al.,}{{McLean}
  et~al.}{2012}]{McLean2012}
{McLean} I.~S.,  et~al., 2012, in Ground-based and Airborne Instrumentation for
  Astronomy IV. p. 84460J, \mn@doi{10.1117/12.924794}

\bibitem[\protect\citeauthoryear{{Momcheva} et~al.,}{{Momcheva}
  et~al.}{2016}]{Momcheva2016}
{Momcheva} I.~G.,  et~al., 2016, \mn@doi [\apjs] {10.3847/0067-0049/225/2/27},
  \href {https://ui.adsabs.harvard.edu/abs/2016ApJS..225...27M} {225, 27}

\bibitem[\protect\citeauthoryear{{Nomoto}, {Tominaga}, {Umeda}, {Kobayashi}  \&
  {Maeda}}{{Nomoto} et~al.}{2006}]{Nomoto2006}
{Nomoto} K.,  {Tominaga} N.,  {Umeda} H.,  {Kobayashi} C.,   {Maeda} K.,  2006,
  \mn@doi [Nuclear Physics A] {10.1016/j.nuclphysa.2006.05.008}, \href
  {https://ui.adsabs.harvard.edu/abs/2006NuPhA.777..424N} {777, 424}

\bibitem[\protect\citeauthoryear{{Oke} et~al.,}{{Oke} et~al.}{1995}]{Oke1995}
{Oke} J.~B.,  et~al., 1995, \mn@doi [\pasp] {10.1086/133562}, \href
  {https://ui.adsabs.harvard.edu/abs/1995PASP..107..375O} {107, 375}

\bibitem[\protect\citeauthoryear{{Papovich}, {Finkelstein}, {Ferguson}, {Lotz}
  \& {Giavalisco}}{{Papovich} et~al.}{2011}]{Papovich2011}
{Papovich} C.,  {Finkelstein} S.~L.,  {Ferguson} H.~C.,  {Lotz} J.~M.,
  {Giavalisco} M.,  2011, \mn@doi [\mnras] {10.1111/j.1365-2966.2010.17965.x},
  \href {https://ui.adsabs.harvard.edu/abs/2011MNRAS.412.1123P} {412, 1123}

\bibitem[\protect\citeauthoryear{{P{\'e}rez-Montero}}{{P{\'e}rez-Montero}}{2014}]{PerezMontero2014}
{P{\'e}rez-Montero} E.,  2014, \mn@doi [\mnras] {10.1093/mnras/stu753}, \href
  {https://ui.adsabs.harvard.edu/abs/2014MNRAS.441.2663P} {441, 2663}

\bibitem[\protect\citeauthoryear{{Pilyugin}, {V{\'\i}lchez}, {Mattsson}  \&
  {Thuan}}{{Pilyugin} et~al.}{2012}]{Pilyugin2012}
{Pilyugin} L.~S.,  {V{\'\i}lchez} J.~M.,  {Mattsson} L.,   {Thuan} T.~X.,
  2012, \mn@doi [\mnras] {10.1111/j.1365-2966.2012.20420.x}, \href
  {https://ui.adsabs.harvard.edu/abs/2012MNRAS.421.1624P} {421, 1624}

\bibitem[\protect\citeauthoryear{{Price} et~al.,}{{Price}
  et~al.}{2020}]{Price2020}
{Price} S.~H.,  et~al., 2020, \mn@doi [\apj] {10.3847/1538-4357/ab7990}, \href
  {https://ui.adsabs.harvard.edu/abs/2020ApJ...894...91P} {894, 91}

\bibitem[\protect\citeauthoryear{{Reddy}, {Pettini}, {Steidel}, {Shapley},
  {Erb}  \& {Law}}{{Reddy} et~al.}{2012}]{Reddy2012b}
{Reddy} N.~A.,  {Pettini} M.,  {Steidel} C.~C.,  {Shapley} A.~E.,  {Erb} D.~K.,
    {Law} D.~R.,  2012, \mn@doi [\apj] {10.1088/0004-637X/754/1/25}, \href
  {https://ui.adsabs.harvard.edu/abs/2012ApJ...754...25R} {754, 25}

\bibitem[\protect\citeauthoryear{{Reddy} et~al.,}{{Reddy}
  et~al.}{2018}]{Reddy2018}
{Reddy} N.~A.,  et~al., 2018, \mn@doi [\apj] {10.3847/1538-4357/aaed1e}, \href
  {https://ui.adsabs.harvard.edu/abs/2018ApJ...869...92R} {869, 92}

\bibitem[\protect\citeauthoryear{{Rix}, {Pettini}, {Leitherer}, {Bresolin},
  {Kudritzki}  \& {Steidel}}{{Rix} et~al.}{2004}]{Rix2004}
{Rix} S.~A.,  {Pettini} M.,  {Leitherer} C.,  {Bresolin} F.,  {Kudritzki}
  R.-P.,   {Steidel} C.~C.,  2004, \mn@doi [\apj] {10.1086/424031}, \href
  {https://ui.adsabs.harvard.edu/abs/2004ApJ...615...98R} {615, 98}

\bibitem[\protect\citeauthoryear{{Runco} et~al.,}{{Runco}
  et~al.}{2020}]{Runco2020}
{Runco} J.~N.,  et~al., 2020, arXiv e-prints, \href
  {https://ui.adsabs.harvard.edu/abs/2020arXiv200804924R} {p. arXiv:2008.04924}

\bibitem[\protect\citeauthoryear{{Sanders} et~al.,}{{Sanders}
  et~al.}{2016a}]{Sanders2016}
{Sanders} R.~L.,  et~al., 2016a, \mn@doi [\apj] {10.3847/0004-637X/816/1/23},
  \href {https://ui.adsabs.harvard.edu/abs/2016ApJ...816...23S} {816, 23}

\bibitem[\protect\citeauthoryear{{Sanders} et~al.,}{{Sanders}
  et~al.}{2016b}]{Sanders2016a}
{Sanders} R.~L.,  et~al., 2016b, \mn@doi [\apj] {10.3847/0004-637X/816/1/23},
  \href {https://ui.adsabs.harvard.edu/abs/2016ApJ...816...23S} {816, 23}

\bibitem[\protect\citeauthoryear{{Sanders} et~al.,}{{Sanders}
  et~al.}{2018}]{Sanders2018}
{Sanders} R.~L.,  et~al., 2018, \mn@doi [\apj] {10.3847/1538-4357/aabcbd},
  \href {https://ui.adsabs.harvard.edu/abs/2018ApJ...858...99S} {858, 99}

\bibitem[\protect\citeauthoryear{{Sanders} et~al.,}{{Sanders}
  et~al.}{2019}]{Sanders2019}
{Sanders} R.~L.,  et~al., 2019, arXiv e-prints, \href
  {https://ui.adsabs.harvard.edu/abs/2019arXiv191013594S} {p. arXiv:1910.13594}

\bibitem[\protect\citeauthoryear{{Sanders} et~al.,}{{Sanders}
  et~al.}{2020}]{Sanders2020}
{Sanders} R.~L.,  et~al., 2020, \mn@doi [\mnras] {10.1093/mnras/stz3032}, \href
  {https://ui.adsabs.harvard.edu/abs/2020MNRAS.491.1427S} {491, 1427}

\bibitem[\protect\citeauthoryear{{Shapley}, {Coil}, {Ma}  \& {Bundy}}{{Shapley}
  et~al.}{2005}]{Shapley2005}
{Shapley} A.~E.,  {Coil} A.~L.,  {Ma} C.-P.,   {Bundy} K.,  2005, \mn@doi
  [\apj] {10.1086/497630}, \href
  {https://ui.adsabs.harvard.edu/abs/2005ApJ...635.1006S} {635, 1006}

\bibitem[\protect\citeauthoryear{{Shapley}, {Steidel}, {Pettini}, {Adelberger}
  \& {Erb}}{{Shapley} et~al.}{2006}]{Shapley2006}
{Shapley} A.~E.,  {Steidel} C.~C.,  {Pettini} M.,  {Adelberger} K.~L.,   {Erb}
  D.~K.,  2006, \mn@doi [\apj] {10.1086/507511}, \href
  {https://ui.adsabs.harvard.edu/abs/2006ApJ...651..688S} {651, 688}

\bibitem[\protect\citeauthoryear{{Shapley} et~al.,}{{Shapley}
  et~al.}{2015}]{Shapley2015}
{Shapley} A.~E.,  et~al., 2015, \mn@doi [\apj] {10.1088/0004-637X/801/2/88},
  \href {https://ui.adsabs.harvard.edu/abs/2015ApJ...801...88S} {801, 88}

\bibitem[\protect\citeauthoryear{{Shapley} et~al.,}{{Shapley}
  et~al.}{2019}]{Shapley2019}
{Shapley} A.~E.,  et~al., 2019, \mn@doi [\apjl] {10.3847/2041-8213/ab385a},
  \href {https://ui.adsabs.harvard.edu/abs/2019ApJ...881L..35S} {881, L35}

\bibitem[\protect\citeauthoryear{{Sim{\'o}n-D{\'\i}az}}{{Sim{\'o}n-D{\'\i}az}}{2010}]{Simondiaz2010}
{Sim{\'o}n-D{\'\i}az} S.,  2010, \mn@doi [\aap] {10.1051/0004-6361/200913120},
  \href {https://ui.adsabs.harvard.edu/abs/2010A&A...510A..22S} {510, A22}

\bibitem[\protect\citeauthoryear{{Sim{\'o}n-D{\'\i}az} \&
  {Stasi{\'n}ska}}{{Sim{\'o}n-D{\'\i}az} \&
  {Stasi{\'n}ska}}{2011}]{Simondiaz2011}
{Sim{\'o}n-D{\'\i}az} S.,  {Stasi{\'n}ska} G.,  2011, \mn@doi [\aap]
  {10.1051/0004-6361/201015512}, \href
  {https://ui.adsabs.harvard.edu/abs/2011A&A...526A..48S} {526, A48}

\bibitem[\protect\citeauthoryear{{Sommariva}, {Mannucci}, {Cresci}, {Maiolino},
  {Marconi}, {Nagao}, {Baroni}  \& {Grazian}}{{Sommariva}
  et~al.}{2012}]{Sommariva2012}
{Sommariva} V.,  {Mannucci} F.,  {Cresci} G.,  {Maiolino} R.,  {Marconi} A.,
  {Nagao} T.,  {Baroni} A.,   {Grazian} A.,  2012, \mn@doi [\aap]
  {10.1051/0004-6361/201118134}, \href
  {https://ui.adsabs.harvard.edu/abs/2012A&A...539A.136S} {539, A136}

\bibitem[\protect\citeauthoryear{{Stanway} \& {Eldridge}}{{Stanway} \&
  {Eldridge}}{2018}]{Stanway2018}
{Stanway} E.~R.,  {Eldridge} J.~J.,  2018, \mn@doi [\mnras]
  {10.1093/mnras/sty1353}, \href
  {https://ui.adsabs.harvard.edu/abs/2018MNRAS.479...75S} {479, 75}

\bibitem[\protect\citeauthoryear{{Stanway}, {Eldridge}  \& {Becker}}{{Stanway}
  et~al.}{2016}]{Stanway2016}
{Stanway} E.~R.,  {Eldridge} J.~J.,   {Becker} G.~D.,  2016, \mn@doi [\mnras]
  {10.1093/mnras/stv2661}, \href
  {https://ui.adsabs.harvard.edu/abs/2016MNRAS.456..485S} {456, 485}

\bibitem[\protect\citeauthoryear{{Steidel} et~al.,}{{Steidel}
  et~al.}{2014}]{Steidel2014}
{Steidel} C.~C.,  et~al., 2014, \mn@doi [\apj] {10.1088/0004-637X/795/2/165},
  \href {https://ui.adsabs.harvard.edu/abs/2014ApJ...795..165S} {795, 165}

\bibitem[\protect\citeauthoryear{{Steidel}, {Strom}, {Pettini}, {Rudie},
  {Reddy}  \& {Trainor}}{{Steidel} et~al.}{2016}]{Steidel2016}
{Steidel} C.~C.,  {Strom} A.~L.,  {Pettini} M.,  {Rudie} G.~C.,  {Reddy} N.~A.,
    {Trainor} R.~F.,  2016, \mn@doi [\apj] {10.3847/0004-637X/826/2/159}, \href
  {https://ui.adsabs.harvard.edu/abs/2016ApJ...826..159S} {826, 159}

\bibitem[\protect\citeauthoryear{{Strom}, {Steidel}, {Rudie}, {Trainor},
  {Pettini}  \& {Reddy}}{{Strom} et~al.}{2017}]{Strom2017}
{Strom} A.~L.,  {Steidel} C.~C.,  {Rudie} G.~C.,  {Trainor} R.~F.,  {Pettini}
  M.,   {Reddy} N.~A.,  2017, \mn@doi [\apj] {10.3847/1538-4357/836/2/164},
  \href {https://ui.adsabs.harvard.edu/abs/2017ApJ...836..164S} {836, 164}

\bibitem[\protect\citeauthoryear{{Strom}, {Steidel}, {Rudie}, {Trainor}  \&
  {Pettini}}{{Strom} et~al.}{2018}]{Strom2018}
{Strom} A.~L.,  {Steidel} C.~C.,  {Rudie} G.~C.,  {Trainor} R.~F.,   {Pettini}
  M.,  2018, \mn@doi [\apj] {10.3847/1538-4357/aae1a5}, \href
  {https://ui.adsabs.harvard.edu/abs/2018ApJ...868..117S} {868, 117}

\bibitem[\protect\citeauthoryear{{Topping} \& {Shull}}{{Topping} \&
  {Shull}}{2015}]{Topping2015}
{Topping} M.~W.,  {Shull} J.~M.,  2015, \mn@doi [\apj]
  {10.1088/0004-637X/800/2/97}, \href
  {https://ui.adsabs.harvard.edu/abs/2015ApJ...800...97T} {800, 97}

\bibitem[\protect\citeauthoryear{{Topping}, {Shapley}, {Reddy}, {Sanders},
  {Coil}, {Kriek}, {Mobasher}  \& {Siana}}{{Topping}
  et~al.}{2019}]{Topping2020}
{Topping} M.~W.,  {Shapley} A.~E.,  {Reddy} N.~A.,  {Sanders} R.~L.,  {Coil}
  A.~L.,  {Kriek} M.,  {Mobasher} B.,   {Siana} B.,  2019, arXiv e-prints,
  \href {https://ui.adsabs.harvard.edu/abs/2019arXiv191210243T} {p.
  arXiv:1912.10243}

\bibitem[\protect\citeauthoryear{{Veilleux} \& {Osterbrock}}{{Veilleux} \&
  {Osterbrock}}{1987}]{Veilleux1987}
{Veilleux} S.,  {Osterbrock} D.~E.,  1987, \mn@doi [\apjs] {10.1086/191166},
  \href {https://ui.adsabs.harvard.edu/abs/1987ApJS...63..295V} {63, 295}

\makeatother
\end{thebibliography}

\end{document}